\documentclass[reprint,superscriptaddress, amsmath,amssymb,aps,longbibliography]{revtex4-2}

 \usepackage[bottom]{footmisc}

\usepackage{color}
\RequirePackage[dvipsnames,usenames]{xcolor}
\usepackage[T1]{fontenc}
\usepackage[latin9]{inputenc}
\usepackage{enumitem}
\usepackage{tikz}
\usetikzlibrary{positioning}
\usepackage{mathrsfs}
\usepackage[maxfloats=256]{morefloats}
\usepackage{bm}
\usepackage{amsmath}
\usepackage{amsthm}
\usepackage{amssymb}
\usepackage{stmaryrd}
\usepackage{hyperref}
\usepackage{mathtools}
\usepackage{dsfont}

\newcommand{\ud}{\mathrm{d}}

\edef\ordinarycolon{\mathchar\the\mathcode`: }
\edef\ordinaryequals{\mathchar\the\mathcode`= }
\DeclareBoldMathCommand\boldlangle{\left\langle}
\DeclareBoldMathCommand\boldrangle{\right\rangle}

\usepackage{breqn}
\catcode`^=7

\AtBeginDocument{%
  \catcode`^=12
}

\makeatletter

\allowdisplaybreaks

\let\cat@comma@active\@empty

\usepackage[capitalize]{cleveref}

\DeclareMathOperator*{\supp}{supp}

\newif\ifnotes
\notestrue

\renewcommand{\mid}{\vert}

\makeatother

\newcommand{\ba}{\begin{eqnarray}}
\newcommand{\ea}{\end{eqnarray}}

\newcommand{\eq}[1]{\begin{align}#1\end{align}}

\newcommand{\tvx}{{\pmb{x}^\dag}}

\newcommand{\vP}{{\bold{P}}}
\newcommand{\ovP}{\overline{\bold{P}}}
\newcommand{\tvP}{{\bold{P}^\dag}}

\newcommand{\do}{\dot{\omega}}

\newcommand{\vx}{{\pmb{x}}}

\begin{document}

\preprint{}

\title{Nonequilibrium thermodynamics of uncertain stochastic processes}

\author{Jan Korbel}
 \affiliation{Section for Science of Complex Systems, CeMSIIS, Medical University of Vienna, Spitalgasse 23, 1090 Vienna, Austria}
\email{jan.korbel@meduniwien.ac.at}
\affiliation{Complexity Science Hub Vienna, Josefst\"{a}dter Strasse 39, 1080 Vienna, Austria}
\author{David H. Wolpert}%
\affiliation{Santa Fe Institute, Santa Fe, NM, USA}
\email{david.h.wolpert@gmail.com}
\homepage{ http://davidwolpert.weebly.com}
\affiliation{Complexity Science Hub Vienna, Josefst\"{a}dter Strasse 39, 1080 Vienna, Austria}
\affiliation{Arizona State University, Tempe, AZ, USA}
\affiliation{International Center for Theoretical Physics, Trieste, Italy}
\affiliation{Albert Einstein Institute for Advanced Study, New York, USA}






\date{\today}

\begin{abstract}
Stochastic thermodynamics is formulated under the assumption of perfect knowledge of all
thermodynamic parameters. However, in any real-world experiment, there is
non-zero uncertainty about the precise value of temperatures, chemical potentials, energy spectrum, etc. Here we investigate
how this uncertainty modifies the theorems of stochastic thermodynamics. We consider two scenarios: in the (called \emph{effective}) scenario we fix the (unknown, randomly generated)
experimental apparatus and then repeatedly observe (stochastic) trajectories of the system
for that fixed apparatus. In contrast, in a (called \emph{phenomenological}) scenario the (unknown) apparatus is
re-generated for each trajectory. We derive expressions for thermodynamic quantities in both scenarios. We also
discuss the physical interpretation of effective (scenario) entropy production (EP), derive the effective mismatch cost,
and provide a numerical analysis of the effective thermodynamics of
a quantum dot implementing bit erasure with
uncertain temperature. We then analyze the protocol for moving between two state distributions that
maximize effective work extraction.
Next, we investigate the effective thermodynamic value of information, focusing on the case where
there is a delay between the initialization of the system and the start
of the protocol. Finally, we derive the detailed and integrated fluctuation theorems (FTs) for the phenomenological EP.
In particular, we show how the phenomenological FTs account for the fact that the longer a trajectory
runs, the more information it provides concerning the precise experimental apparatus, and therefore the
less EP it generates.
\end{abstract}

\maketitle

\section{Introduction}
The microscopic laws of classical and quantum physics are parameterized sets of equations
that specify the evolution of a closed system starting from a specific state. To use those equations, we
need to know that specific state, we need to be sure the system is closed, and we need to know the
values of the parameters in the equations~\cite{seifert2012stochastic,van2015ensemble}.

Unfortunately, in many real-world scenarios, we are uncertain about the precise state of the system, and
very often, the system is open rather than closed, subject to
uncertain interactions with the external environment.
Statistical physics accounts for these two types of uncertainty by building on the microscopic laws of physics in two ways.
First, to capture uncertainty about the state of the system,
we replace the exact specification of the system's state with a probability distribution
over states. Second, to capture uncertain interactions between the system and the external environment, we add randomness to the
dynamics in a precisely parameterized form~\footnote{We note though there is a substantial
literature which adopts the ``inclusive'' framework, in which the external environment is finite, and
the joint dynamics of the system-environment is explicitly
modeled. This framework is the classical form of ``open quantum thermodynamics''.
The inclusive framework  has been explored both for explicit Hamiltonian dynamics
over the joint system, where the only randomness  in the initial
state~\cite{kawai_dissipation:_2007,esposito2010entropy,ptaszynski2018first}, and for approximate Hamiltonian
dynamics~\cite{seifert2016first,strasberg2017stochastic,talkner2020HMF,talkner2020comment,strasberg2020measurability}.}.

In particular, in the sub-field of classical stochastic thermodynamics~\cite{seifert2012stochastic,van2015ensemble}
we model the system as a probability distribution evolving under a continuous-time Markov chain (CTMC)
with a precisely specified rate matrix. Often in this work, we require that the CTMC obeys local detailed balance (LDB). This
means that the rate matrix of the CTMC has to obey certain restrictions, which are parameterized by the energy
spectrum of the system, the number of thermodynamic reservoirs in the external environment perturbing
the system's dynamics, and the temperatures and chemical potentials of those reservoirs. Often we also
allow both the Hamiltonian of the system and the rate matrix of the associated
CTMC to change in time in a deterministic manner, perhaps coupled by LDB.
That joint trajectory is referred to as a  ``protocol''. 

However, in addition to uncertainty about the state of the system and uncertainty about interactions
with the external environment, there is an additional unavoidable type of uncertainty in all real-world systems: uncertainty
about the parameters in the equations governing the dynamics. In the context of stochastic thermodynamics, this
means that even if we impose LDB, we will \textit{never} know the reservoir temperatures and
chemical potentials to infinite precision (often even being unsure about the number of such reservoirs),
we will never know the energy spectrum to infinite precision, and more generally, we will never know the rate matrix
and its time dependence to infinite precision.

At present, almost nothing is known about the thermodynamic consequences of this third type of uncertainty despite
its unavoidability~\footnote{We note though that it is
already known that if we do not account for all thermodynamic reservoirs, we invariably under-estimate the total
entropy production in a process~\cite{esposito2010three}.}.
In this paper, we start to fill in this gap by considering how stochastic thermodynamics (and non-equilibrium statistical physics more
generally) needs to be modified to account for this third type of uncertainty, in addition to the two types of uncertainty it already
captures.

We define an \emph{apparatus} $\alpha \in A$ to be any specific set of values of the thermodynamic parameters
of an experiment, including the number of reservoirs, their temperatures and chemical potentials, 
the precise initial distribution over states (i.e., how the system was prepared)
the (deterministic trajectories of the) rate matrices, the (deterministic trajectories of the) energy functions, etc.
Here and throughout, we assume that
these thermodynamic parameters are appropriately related by LDB for any specific $\alpha$.
For simplicity, we also assume that for all apparatuses, the system has the same state space, $X$.
Also for simplicity, we assume that all non-protocol components of an apparatus (in particular the temperatures
and chemical potentials) do not change in time.
In addition, we assume that for all $\alpha$, the process takes place in
the same time interval, $[t_i, t_f]$.
We write an element of $X$ as $x$, and a trajectory of $X$ values across $[t_i, t_f]$ as $\vx$.

We suppose that $\alpha$ is not precisely known and write its probability measure as $\ud P^\alpha$.
Physically, it may be that we have an infinite set of apparatuses
generated by IID sampling $\ud P^{\alpha}$. Alternatively, $\ud P^{\alpha}$ could represent Bayesian
uncertainty or a detailed model of the noise in the measuring instruments used to set the parameters in $\alpha$.
(Below, we will often abuse notation/terminology and refer to a ``distribution'' over apparatuses when
properly speaking, we should be couching the discussion in terms of a probability measure.)
Abusing notation, we will use $A$ to denote both the random variable with values $\alpha$,
and the event space of that random variable \footnote{Note
that $\alpha$ is a generic characterization of apparatuses, which can be formalized as a vector with some components
finite-valued, some countable, and some uncountable, with the measure $\ud P^\alpha$ defined appropriately.}.

Concretely, we consider two kinds of experimental
scenarios. Both start by sampling $\ud P^{\alpha}$, but they differ after that:
\begin{enumerate}[label= \Roman*)]
\item
In the effective scenario, we generate an apparatus by sampling $\ud P^\alpha$. For that fixed apparatus we
then generate
many stochastic trajectories $\vx$.
After running all those trajectories for that fixed apparatus, we can, if we wish, rerun the scenario,
generating another sample of the distribution over apparatuses, which we then use to generate a new
set of stochastic trajectories. We call this the \emph{effective} scenario.

Experimentally, in the effective scenario, one can generate and then observe frequency counts of the distribution $p(\vx | \alpha)$ for multiple random values of $\alpha$, but without ever directly observing $\alpha$.
For example, the experimenter might construct a bit-eraser experimental apparatus involving
a single thermal reservoir whose temperature is fixed throughout the experiment but only
known to the finite precision of $.1^{\circ} K$. The experimenter then runs their experiment many times using this fixed apparatus and collected statistics concerning the trajectories
across those experiments. They can then use those estimates to make (perhaps Bayesian) estimates
of thermodynamic functions of a trajectory, like the associated entropy production.

\item In the phenomenological scenario, we again generate an apparatus by sampling $\ud P^\alpha$, but the apparatus cannot be
fixed while we generate multiple trajectories. Instead, in order to generate a new trajectory we must first generate a new
apparatus by resampling $\ud P^\alpha$. We call this the \emph{phenomenological} scenario.

To illustrate the phenomenological scenario we can return to the example where the experimenter constructs a bit-eraser experimental apparatus involving
a single thermal reservoir whose temperature at the beginning of the experiment is only
known to some finite precision of $.001^{\circ} K$. Suppose though that the temperature is very
slowly drifting randomly in time. Any given run of the experiment is very fast on the timescale
of that drift, so we can treat the temperature as fixed throughout the run. However, after generating a trajectory
by running the experiment, it takes a long time for the system to be reinitialized to rerun the experiment, and during
that time the temperature has drifted to a new value that is statistically independent of the value
during the preceding run. As in the effective scenario, the experimenter runs their experiment
many times and collected statistics concerning (functions of) the trajectories
across those experiments. However, in the phenomenological scenario, one can only observe
frequency counts of the $\alpha$-averaged distribution over trajectories,
$\bar{p}(\vx) := \int \ud P^\alpha p(\vx | \alpha)$.
\end{enumerate}


\begin{figure*}
\centering
\includegraphics[width=0.4\linewidth]{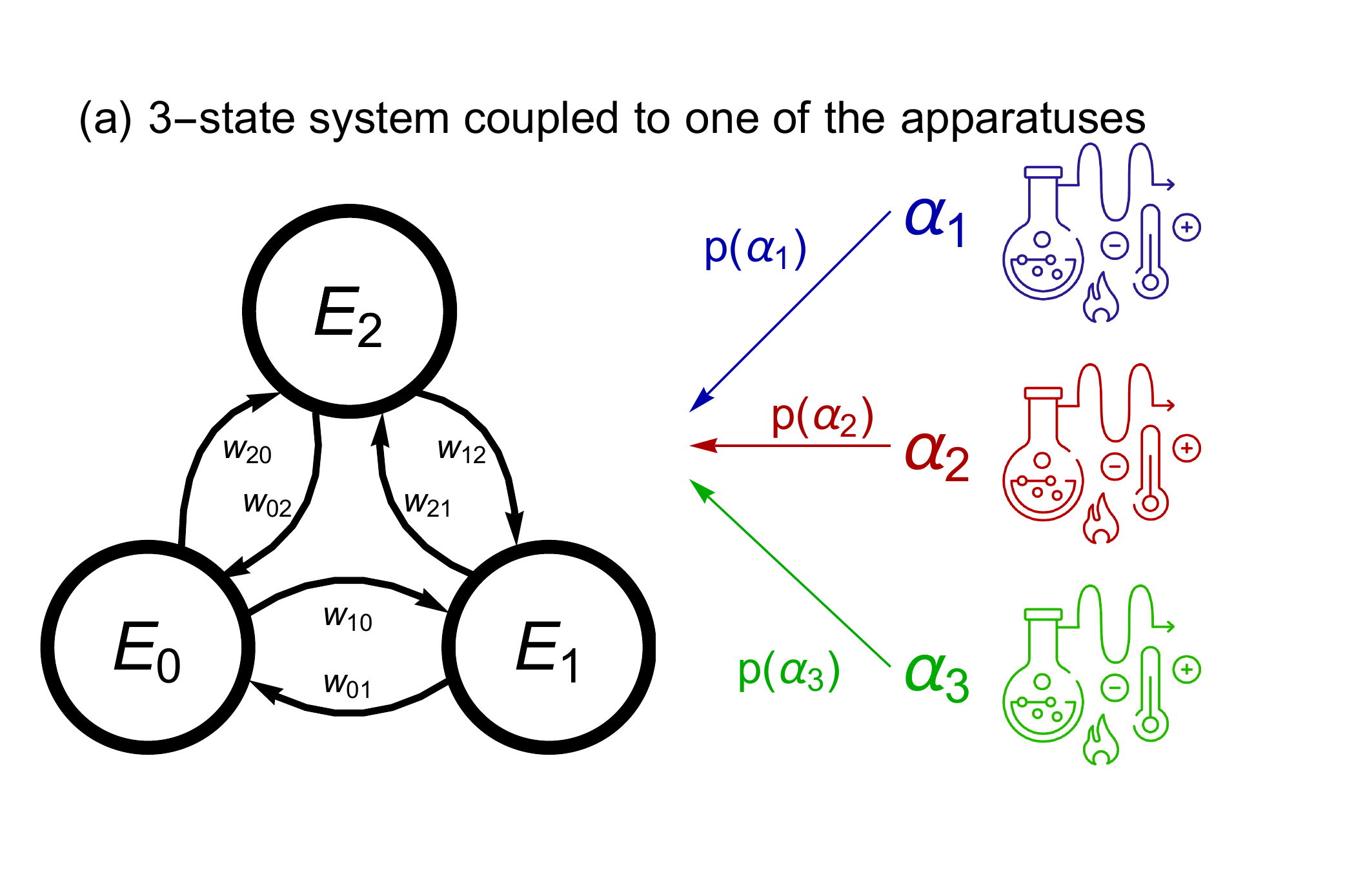}
\includegraphics[width=0.4\linewidth]{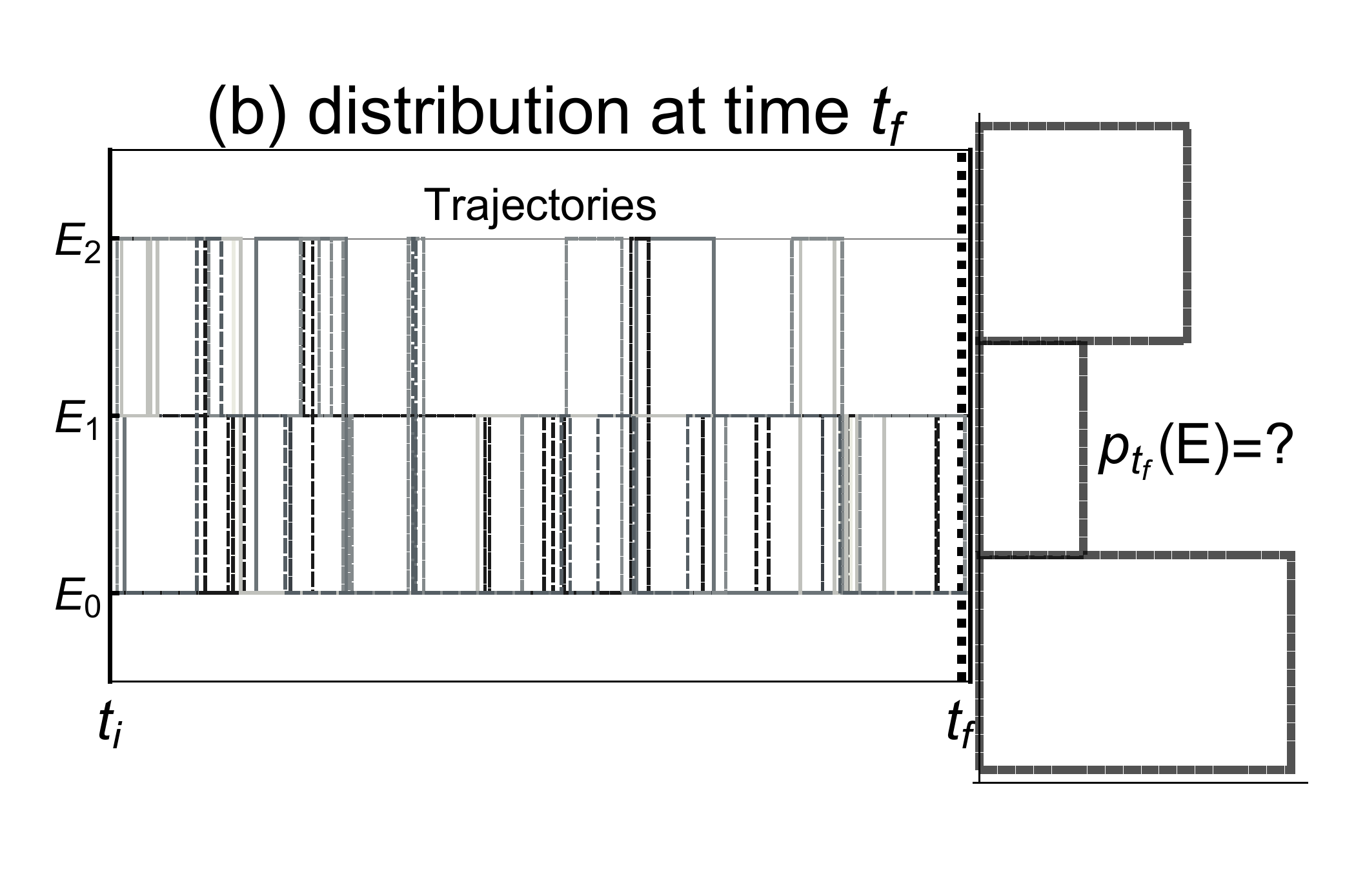}\\
\includegraphics[width=0.4\linewidth]{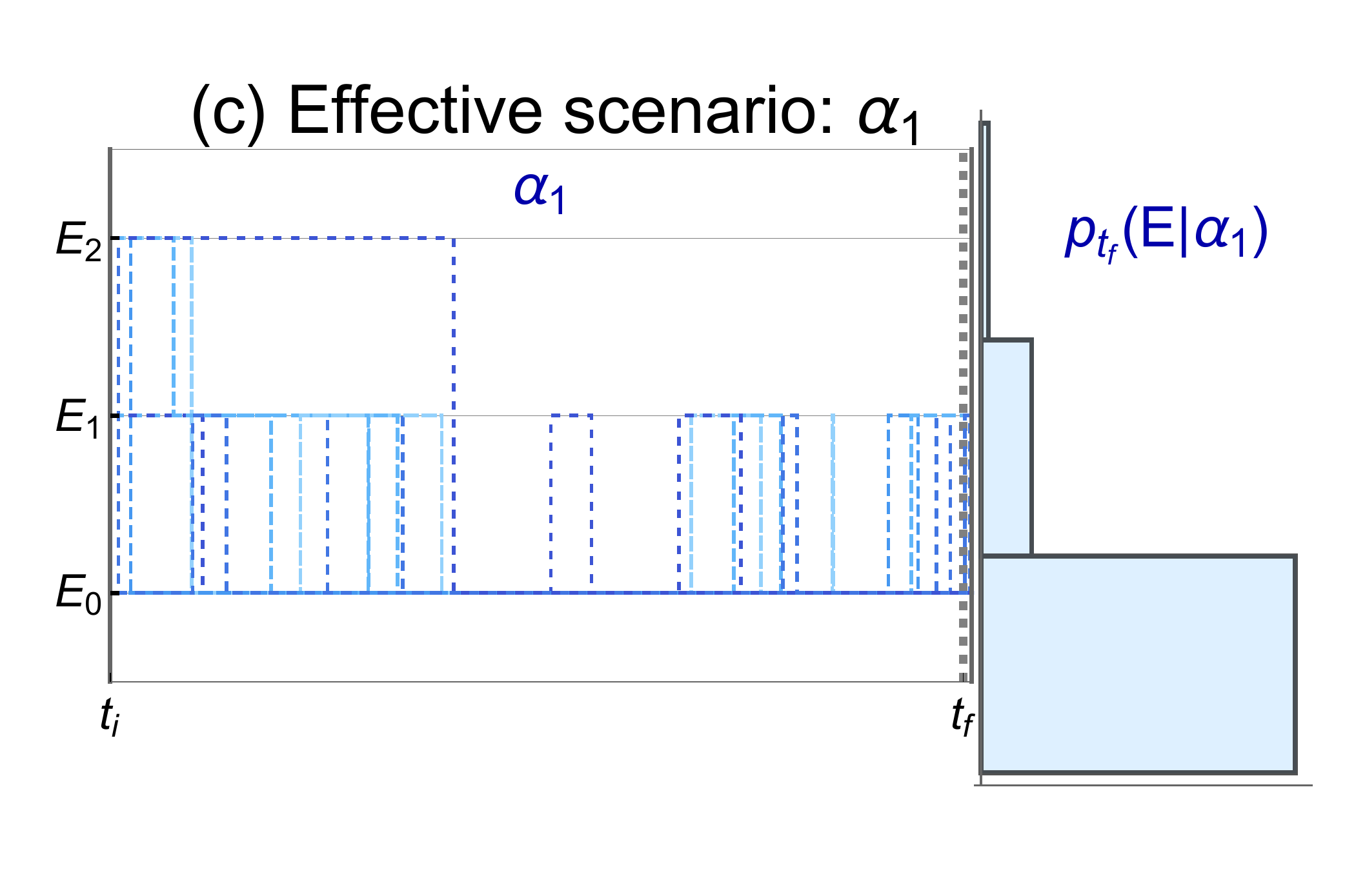}
\includegraphics[width=0.4\linewidth]{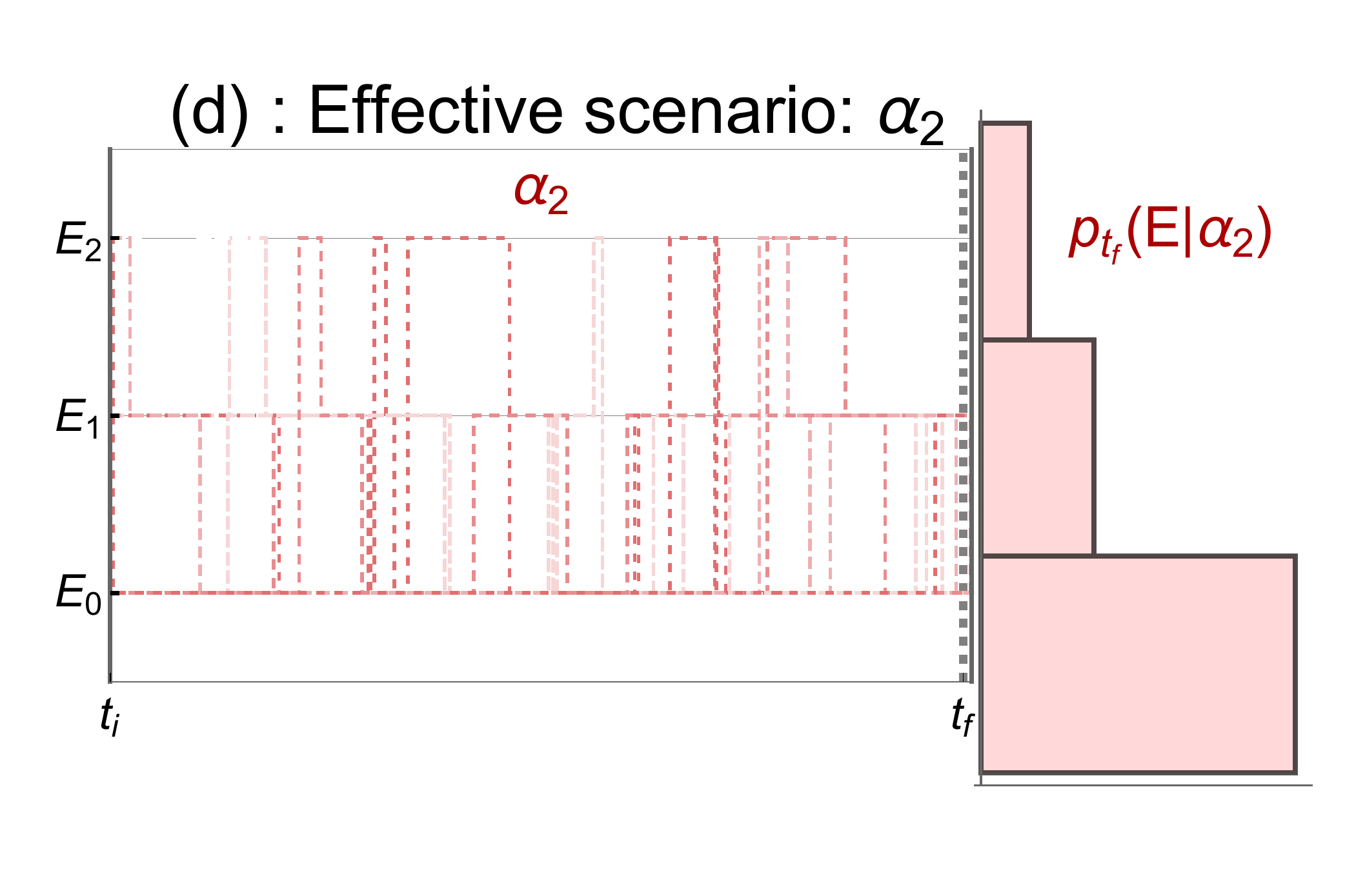}\\
\includegraphics[width=0.4\linewidth]{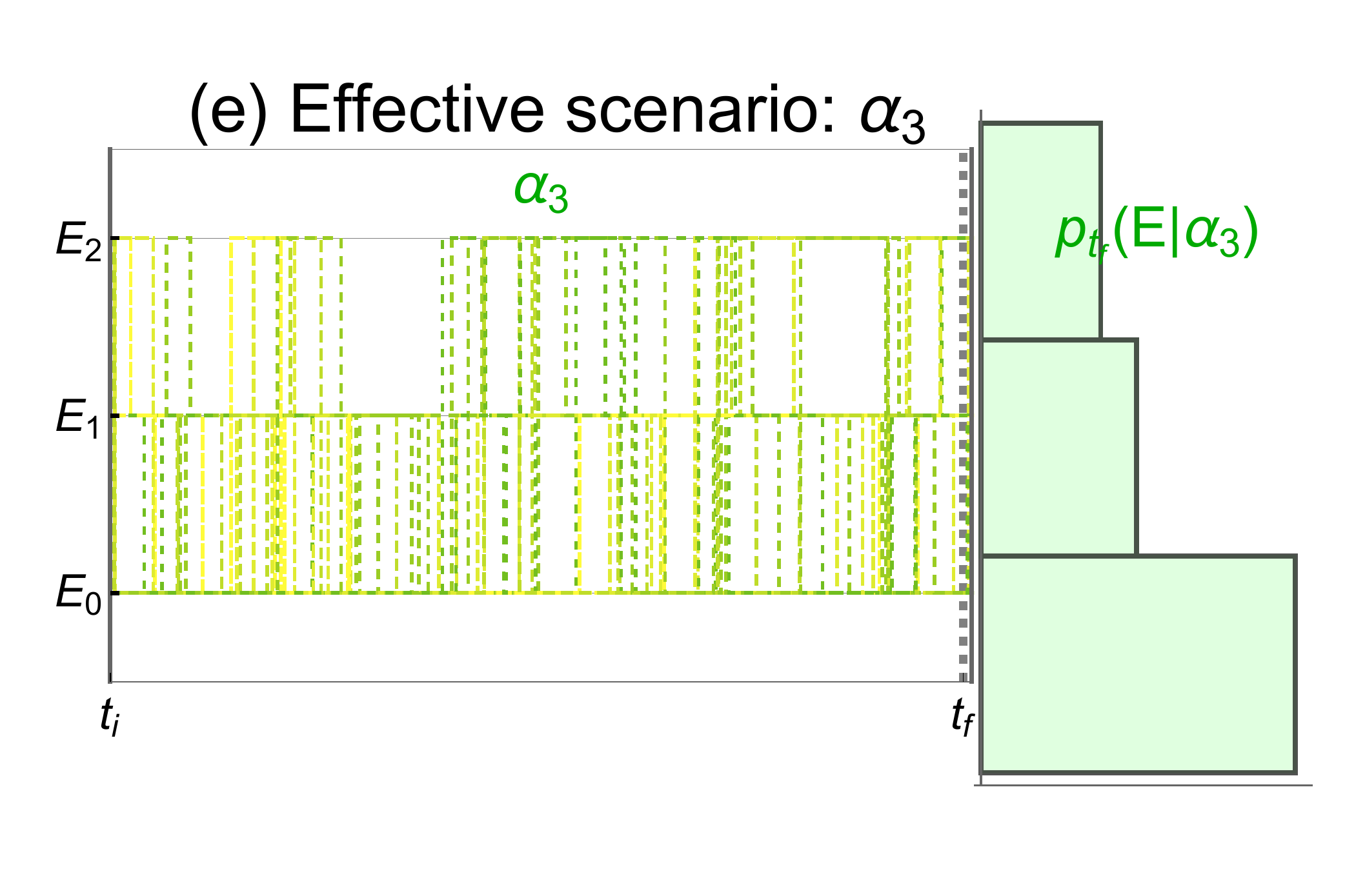}
\includegraphics[width=0.4\linewidth]{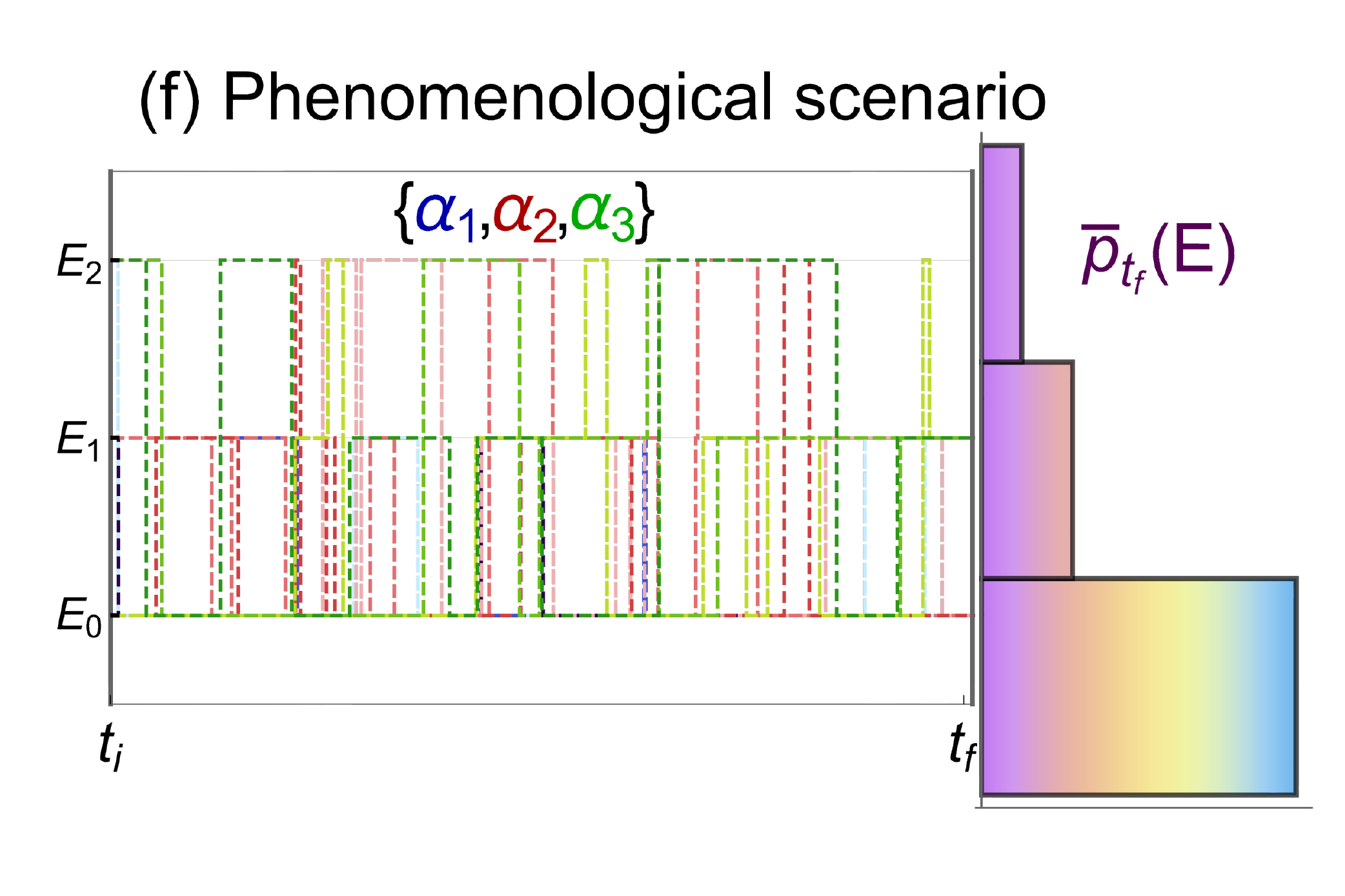}
\caption{Comparison of the two uncertain apparatus scenarios considered in this paper. a) A simple,
a three-state system that can be coupled to one of the three apparatuses, with respective probabilities. b)
We will plot trajectories of the values of a quantity $E$ that has three possible values across the time interval $[t_i, t_f]$, along
with the associated empirical estimate of the relative probabilities that the system had each of those three values
during $[t_i, t_f]$
c),d),e)
Plots for the effective scenario, where we can estimate the marginal distribution at time $t_f$
$p_{t_f}(E|\alpha)$ for a fixed apparatus $\alpha$, shown for three separate instances of the scenario corresponding to the
three possible apparatuses.
f) The phenomenological scenario, in which each trajectory
is sampled with a different apparatus, and therefore only $\bar{p}_{t_f}(E)$ can be estimated.
}
\label{fig:1}
\end{figure*}
\noindent
Illustrations of both scenarios are depicted in Fig. \ref{fig:1},
for a simple three-state time-homogeneous system coupled to one of the three possible apparatuses. In each case, we measure the marginal probability distribution at the final time $t_f$.
Note that in neither scenario do we allow any \textit{direct} measurement of $\alpha$.
However, there may be indirect information about $\alpha$ that arises from the precise trajectory of
states that is generated once $\alpha$ is chosen.

Crucially, the ensemble-level thermodynamic quantities generated in these two scenarios can differ, since
the two types of average involved (once over $\alpha$, once over $x$) do not necessarily commute.
As an example, in the effective scenario, since we can form an estimate of $p_t(x \mid \alpha)$
by running many iterations of a fixed experimental apparatus, we can experimentally estimate the
entropy defined as
\eq{
	\bar{S}(P_t) = -\int \ud P^\alpha \sum_x p_t(x | \alpha) \ln p_t(x | \alpha)
\label{eq:1}
}
using empirical frequency counts.

This is not possible in the phenomenological scenario, in which we can only experimentally estimate a more ``coarse-grained'' version of entropy,
\eq{
	S(\bar{P}_t) = - \sum_x \left(\int \ud P^\alpha p_t(x | \alpha)\right) \ln \left(\int \ud P^\alpha p_t(x | \alpha)\right)
\label{eq:2}
}
using empirical frequency counts.

Note also that if we are in the effective scenario and have the ability
to force a new apparatus to be (randomly) generated whenever we want, then we
can implement the phenomenological scenario just by forcing a new apparatus to be generated after every run. In this
augmented version of the effective scenario, we could experimentally estimate the quantity in \cref{eq:2}
using empirical frequency counts. However,
if we are in the effective scenario and do not have this extra ability, then we cannot estimate the quantity
in \cref{eq:2}, only the quantity in \cref{eq:1}. (In this paper,
whenever we discuss the effective scenario, we will assume we do not have this extra ability.)

The difference between the thermodynamics of the two scenarios will be a central focus of our analysis below.

\subsection*{Related research}
It is important to distinguish between the focus of this paper and some of the issues that
have been investigated in the recent literature. Some recent research has considered
how to modify stochastic thermodynamics if the experimentalist is not able to view all state transitions in the system
as it
evolves~\cite{bisker2017hierarchical,shiraishi_ito_sagawa_thermo_of_time_separation.2015}.
The uncertainty in these papers concerns what is observed as the system evolves, whereas we focus on
uncertainty in the parameters governing
that evolution. Similarly, some models consider either spatial \cite{matsuo2000stochastic} or temporal \cite{fiore2019entropy}
variation of temperature and other parameters, but they assume that this evolution is known.
In contrast, we assume that $\alpha$ is fixed throughout the interval, but to an unknown value.

Probably the closest research to what we consider in this paper is sometimes called \emph{superstatistics}.
It has long been known that an average over Gibbs distributions cannot be written
as some single Gibbs distribution (Thm.\,1 in~\cite{wolpert1993use}). This means that even equilibrium statistical physics
must be modified when there is uncertainty in the temperature of a system.
The analysis of these modifications was begun by Beck and Cohen \cite{beck2003superstaitstic},
who developed an  \emph{effective theory} for thermodynamics with temperature fluctuating in time. They considered a system coupled to a
bath,
which is in a \emph{local equilibrium} under the slow evolution of the temperature of the bath.
The main assumption they exploit is \emph{scale-separation}: while for short time scales, the distribution over states
of the system is an equilibrium, canonical
distribution with inverse temperature $\beta$, the long-scale behavior is determined by a \emph{superposition} of canonical
distributions with some distribution of temperatures $f(\beta)$.
The resulting \emph{superstatistical} distribution $p(E) = \int \mathrm{d} \beta f(\beta) \exp(-\beta E)/Z(\beta)$ was later
identified with the distribution corresponding to generalized entropic functionals
\cite{hanel2011generalized,garcia2011superstatistics}
due to the fact that particular generalized entropic functionals are maximized by the same distribution that can be obtained the
superposition of the canonical distribution
with given $f(\beta)$ \cite{BECK200696,mathai2007pathway}.

Later interpretations of superstatistics are not based on the notion of
local equilibria but rather on the
Bayesian approach to systems with uncertain temperature \cite{sattin2006bayesian,davis2018temperature}. These are
conceptually closer to the focus of this paper, which focuses on off-equilibrium systems that are evolving quickly on the scale of the coupling with the thermal reservoirs, and so cannot
be modeled in terms of time-scale separation.

Similar to the quasi-equilibrium scenarios considered in superstatistics, other
research has focused on
deriving an \emph{effective} description of the system in local equilibrium averaged over
uncertain thermodynamic parameters.
In particular, this is the basis of a very rich and well-studied approach to
analyzing {spin glasses} \cite{edwards1975theory,PhysRevLett.35.1792}, in which the coupling constants $J_{ij}$ in the spin-glass
Hamiltonian  $H = -\sum_{(ij)} J_{ij} s_i s_j$,
are random variables drawn from a given distribution $p(J_{ij})$. Given such a distribution,
the famous replica trick $\overline{\ln Z} = \lim_{n \to 0} \frac{Z^n-1}{n}$ \cite{mezard1987spin}
can be used to calculate the Helmholtz free energy, averaged over all $J_{ij}$. Let us note that in the terminology used in disordered systems, the annealed disorder corresponds to the effective scenario while quenched disorder corresponds to the phenomenological scenario.


Finally, several authors
\cite{kolchinsky2016dependence,wolpert_thermo_comp_review_2019,wolpert_kolchinsk_first_circuits_published.2020,riechers2021initial} investigated the case
where the initial distribution differs from the one that would minimize EP. It describes the situation when the system is designed
by a scientist who was mistaken in their assumption concerning the initial distribution. In this case, the choice of the non-optimal solution generates the extra entropy production that can be described by the so-called mismatch cost.
These papers do not involve a distribution different initial distribution that is re-sampled each time the experiment is re-run.

\subsection*{Roadmap}

One of the major themes of our investigation is that some of the details of how an experiment is conducted
that experimenters currently do not consider in fact have major effects on the precise forms of various thermodynamic
quantities. This is reflected in the difference between (the thermodynamics of) the effective and phenomenological
scenarios, discussed above. Even within the effective scenario though, there are some
important distinctions between different ways of running the experiment (and so different ways of defining
thermodynamic quantities). In particular, there is a major effect on the thermodynamics itself that arises from whether the experimenter
the protocol (time-dependent trajectory of Hamiltonians of the system)
changes from one run of an experiment to the next, or instead is fixed in all runs.
We call these the ``unadapted'' and ``adapted'' situations, respectively.
We start in \cref{sec:ill_example} 
with a simple illustrative example of these two situations, involving
a moving optical tweezer with uncertain stiffness parameter. 

We then  begin our more general analysis. 
First, in \cref{sec:2}, we introduce the necessary notation and briefly recall the main
results of traditional, full-certainty stochastic thermodynamics. In \cref{sec:3}, we present the general form that
stochastic thermodynamics takes in the effective scenario (recall the discussion of the effective and
phenomenological scenarios in the introduction).
We begin by noting that the evolution of the effective probability distribution is not Markovian. Then we derive the
forms of the first and second laws of thermodynamics for effective thermodynamic quantities. Next we
discuss
the relation between effective EP and effective dissipated work.
We illustrate this discussion
 with the numerical example of a fermionic bit erasure with uncertain temperature.
We end this
section by investigating the special case
where the only uncertainty concerns the initial distribution, calculating the associated effective mismatch cost.

In \cref{sec:4}, we focus on (feedback) control protocols for uncertain apparatuses. In contrast to the conventional case where the apparatus is precisely
known, we assume we cannot tailor the protocol for each (uncertain) apparatus separately, but instead
must use the same protocol for all apparatuses. We use this setting to investigate how apparatus
uncertainty affects a foundational concern of stochastic thermodynamics: How much work
can be extracted from a system during a process that takes it from a given initial distribution to a given
target distribution.

First, we consider this issue when we are uncertain both about the
initial distribution (though not the final one) and about the temperature of the system as it evolves.
We focus on how that uncertainty changes the results of
the standard analysis of this issue, in which we suppose a \{quench; equilibrate; semi-statically-evolve\} process is applied
to the system immediately after the initial distribution is generated.

Next, we use this analysis to consider how uncertainty affects the ``thermodynamic value of information''
to a feedback controller~\cite{cao_thermodynamics_2009,sagawa_nonequilibrium_2012}. We restrict attention to
the special case of the analysis where the temperature is known exactly, so
the only uncertainty is in
the initial distribution. We also suppose that there is a (perfectly known) delay between when the initial distribution is
generated, $t_i$, and
the time $\tau$ when the \{quench; equilibrate; semi-statically-evolve\} process can begin, during which time the system evolves
according
to a (perfectly known) rate matrix. In particular, we derive expressions for how the thermodynamic value of
information varies with the length of the delay.

In \cref{sec:5}, we investigate the ensemble entropy production calculated from effective trajectory
probabilities, i.e., from trajectory probabilities given by averaging over apparatuses.
We call this the \emph{phenomenological (ensemble) EP}. We begin by proving that phenomenological
ensemble EP is a lower bound on the average over apparatuses of the
effective ensemble EP. So fixing the apparatus and averaging over the trajectories
--- though without knowing what value the apparatus
is fixed to --- and then averaging over apparatuses increases EP, compared to the case where we average apparatuses
before averaging trajectories.

The difference between effective EP and phenomenological EP is called \emph{likelihood EP}. It measures the difference between
log-likelihood functions
estimated from the forward and time-reversed trajectories.

Considering trajectory versions of all three EPs, we
establish three detailed fluctuation theorems (DFT). In
addition to the well-studied DFT in the literature which concerns a single, known apparatus, we establish the DFT for the
phenomenological EP and for likelihood EP.
The former represents the effective irreversibility of the system by coarse-graining all the apparatuses. The latter represents how
irreversibility affects the estimation of the apparatus' parameters when estimated by observing the forward and time-reversed
trajectories. These results are illustrated by a simple example of a two-state system coupled to one heat reservoir with uncertain temperature.

The paper ends with a discussion section in which we describe just a few of the myriad directions for future work.

\section{Illustrative example}
\label{sec:ill_example}

In this section we illustrate the importance of accounting for the uncertainty of the system parameters in an experiment, 
with a simple example of a colloidal particle in a moving laser trap.
The dynamics of the particle is given by the overdamped Langevin equation
$$ \dot{x} = - \mu \frac{\partial V}{\partial x} + \xi$$
where $\xi$ is the white noise, and $V$ is the potential. Let us consider that the particle is dragged by an optical tweezer with the harmonic potential
$$V_k(x,t) = \frac{k}{2}(x-\lambda(t))^2$$
where $k$ is the stiffness parameter and $\lambda(t)$ is the control protocol. The average work is given by the Sekimoto formula
$$W[\lambda(t)] = \int_{0}^{t_f} \mathrm{d} t \dot{\lambda} \left\langle \frac{V_k(\lambda(t),x(t))}{\partial \lambda} \right\rangle$$
where $\langle .. \rangle$ is the ensemble average. Let us consider $\mu=1$.
Our aim is to move the trap from $\lambda_i = 0$ at time $t_i = 0$ to $\lambda_f$ at time $t_f$ such that the average work is minimal. Following \cite{Udo}, it is possible to express the optimal protocol starting that minimizes the average work as
$$\lambda_k^\star = \frac{\lambda_f(1+kt)}{2+kt_f}$$
and the corresponding optimal work as
\eq{
W_k^\star = \frac{k \lambda_f^2}{2+kt_f}
\label{eq:baby}
}
The complete derivation is done in \cref{app:tweezer}.

We focus on the realistic situation where the experimenter has to measure the stiffness parameter to be able to determine the optimal protocol. The estimation is typically done by repeated measurement of $k$, which leads to a histogram of $k$. 
In practice, often an experimenter will implicitly assume that the uncertainty in $k$ is due to the measurement and takes the average value of stiffness $\bar{k}$ as the single possible value. However, often the uncertainty in the parameters can have a physical reason, e.g., imprecise calibration of the laser. In those kinds of scenarios the stiffness can change for each run of the experiment,
and so the experimenter's implicit assumption is invalid.

Write the stiffness parameter that the experimenter uses to set up the control protocol as $k$, with the real stiffness parameter
written as  $\kappa$ (which in general differs from $k$). In \cref{app:tweezer}
we show that the work can then be expressed as
\begin{eqnarray*}
W_\kappa[\lambda_k(t)] &=& W^\star_k \nonumber\\
&&+ \frac{\lambda_f^2}{(2+kt_f)^2} \left(\frac{\kappa^2-k^2}{\kappa} + \frac{(k-\kappa)^2}{\kappa} e^{-\kappa t_f} \right)
\end{eqnarray*}
using the definition in \cref{eq:baby}.

Let us now assume that the experimenter repeats the experiment many times. In each 
run, the stiffness parameter $\kappa$ is drawn from a distribution $p(\kappa)$. We want to compare two 
broad types of scenario. In the first type of scenario, the
experimenter re-measures the stiffness parameter before each
run of the experiment, so that they can adapt their protocol in that run to the new value of the stiffness parameter
in order to minimizes the average work for that run. This
is an example of what we call an \emph{adapted} scenario below, in \cref{sec:diss_work_types}.  The average work in
this scenario can be expressed as
$$W_{ad.}  = \int \mathrm{d} \kappa W_{\kappa}[\lambda_\kappa(t)]$$

In the second scenario, the experimenter ignores the uncertainty in stiffness, and for each run, 
they assume that the system is described by the same stiffness parameter $\bar{\kappa} = \int \mathrm{d} \kappa p(\kappa) \kappa$. This
is an example of what we call an \emph{unadapted} scenario below, in  \cref{sec:diss_work_types}. 
The minimal expected work in this unadapted scenario can be expressed as
$$W_{unad.} =  \int \mathrm{d} \kappa W_{\kappa}[\lambda_{\overline{\kappa}}(t)]$$

\begin{figure}
\includegraphics[width=0.8\linewidth]{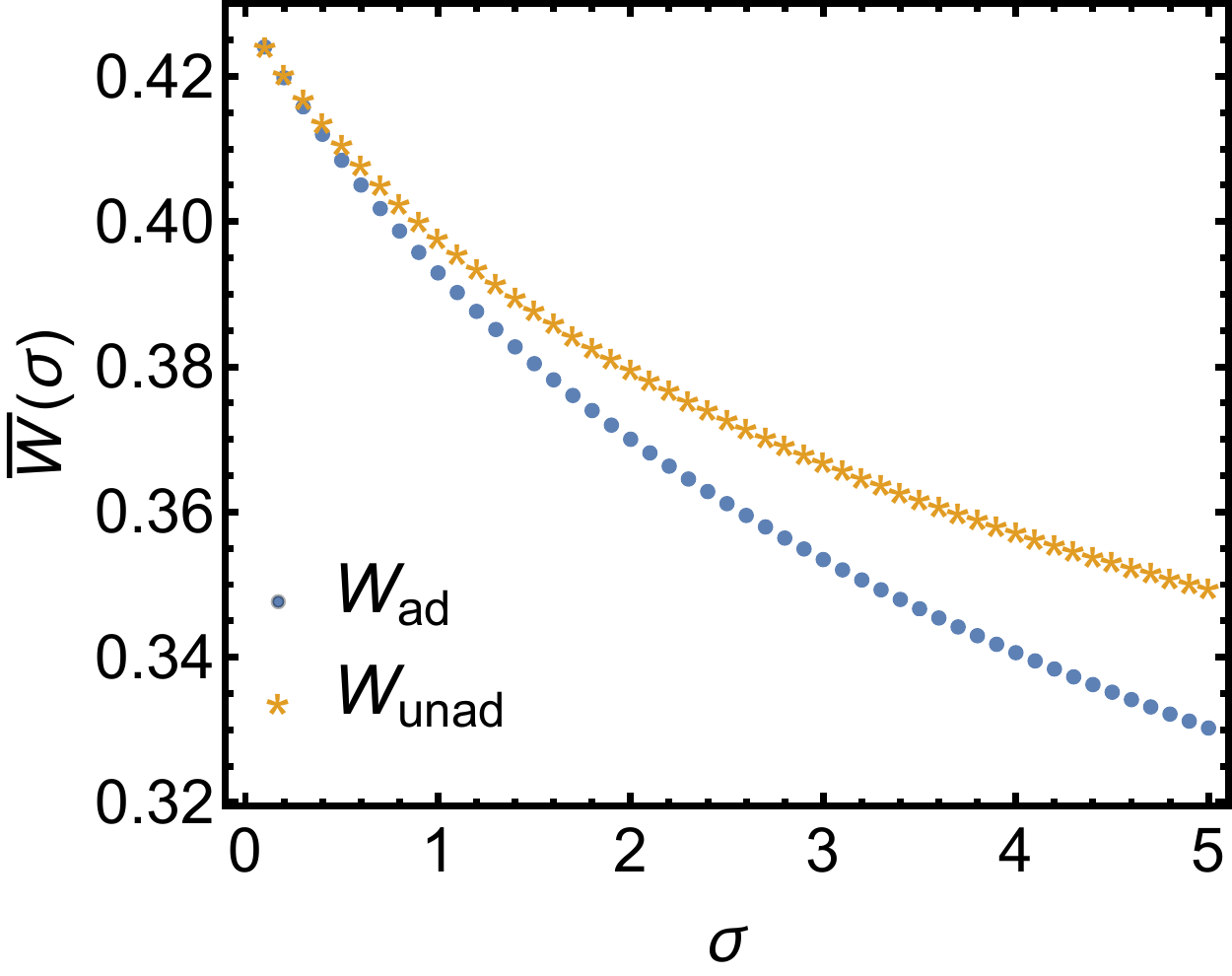}
\caption{The comparison of the average work for an unadapted and adapted scenario for the moving optical tweezer with uncertain stiffness.}
\label{fig:trap}
\end{figure}

Recall that the ``dissipated work'' of a particular protocol is the difference between the work
it expended and the minimal work that would have been expended by an optimal protocol.
Dissipated work is one of the central concerns in non-equilibrium statistical physics. Because
the ``minimal work'' differs between the adapted and unadapted scenarios, we would expect
that the dissipated work does as well. We illustrate this for the current case of a particle
in a trap in \cref{fig:trap}. To generate that figure we chose
$p(\kappa) \sim LogNormal(\mu,\sigma)$, where $\mu$ and $\sigma$ are the mean and the variance of the log-normal distribution~\footnote{N.b., this
is in contrast with what is arguably the most common parametrization, where the parameters are the mean and the variance of the corresponding normal distribution. The transformation between these two parametrizations is given in \cref{app:tweezer}.}. 
The parameters of the lognormal are $\lambda_f = 1$, $t_f = 1$, $\mu \equiv \bar{\kappa} = \frac{3}{2}$. \cref{fig:trap} 
demonstrates the general fact that the adapted and unadapted scenarios can differ in their thermodynamics; with the increasing variance of the distribution, the difference is more pronounced (Note that both adapted and unadapted work decrease with increasing $\sigma$ which is caused by the fact that the expected work is in the case a concave function of stiffness).

Note that choosing the protocol $\lambda_{\overline{\kappa}}(t)$ in this scenario is not the optimal choice that minimizes the average work. (The issue of  protocols that optimize the average work are discussed in \cref{sec:4}.)

\section{Preliminaries}
\label{sec:2}

For simplicity, throughout this paper, we assume the system of interest has a countable state space $X$ with elements
generically written as $x \in X$. A trajectory of values of $X$ across some non-infinitesimal
time interval $[0, t]$ will be written as $\vx_t$, or just $\vx$ for short. The trajectory's state at a specific time $t$ is denoted
as $\vx(t)$.
Throughout this paper, we will often leave the time $t$ implicit.
The Kronecker delta function with arguments $x, x'$ is written as $\delta_{x, x'}$ and
$D_x(p || q) = \sum_x p(x) \ln \frac{p(x)}{q(x)}$ is the Kullback-Leibler (KL) divergence between distributions $p$ and $q$, also
known as the relative entropy~\cite{cover_elements_2012}. When the KL divergence is also averaged over apparatuses, we denote it as
$D_{x,\alpha}(p || q) =  \int \ud P^\alpha \sum_x p(x|\alpha) \ln \frac{p(x|\alpha)}{q(x|\alpha)}$. Throughout this paper, capital
letters indicate random variables that are averaged over, while
lowercase letters indicate specific values of such random variables.

\subsection{Instance-level, trajectory-level and ensemble-level quantities}


A generic function $y_t(x)$ depending on state $x$ (and possibly on time $t$) is called an \emph{instance-level} value.
Instance-level values are denoted by lowercase letters.
Given a distribution over the states $p_t(x)$, we write
\eq{Y_t := \langle y_t \rangle = \int \mathrm{d}x \, p_t(x) y_t(x)}
for the expectation of the instance value $y_t(x)$ under a distribution $p_t(x)$.
We refer to such expected values specified by a particular $t$ as \emph{ensemble}
values, and denote them in uppercase letters.
Also, we will often leave both the distribution and the time $t$ implicit when they are obvious from context.

We refer to quantities specified in terms of
a sequence  $\vx$ of instance values over a time interval $[t_0, t]$ as \emph{trajectory-level},
typically denoted as $\pmb{y}_{[t_0,t]}(\pmb{x})$. When the initial time $t_0$ is fixed, we use only $\pmb{y}_{t}(\pmb{x})$.
Trajectories of values are denoted by bold lowercase letters, with an index $t$ indicating the ending time
of that trajectory.
Given a distribution over trajectories $\bold{P}(\vx)$, we denote the expected value of an associated
sequence as the \textit{ensemble(-trajectory)} value
\eq{\pmb{Y}_t :=  \boldlangle \pmb{y}_t \boldrangle = \int_{t_0}^t \mathcal{D} \vx \, \bold{P}(\vx) \pmb{y}_t(\pmb{x}).} 
where $\mathcal{D} \vx$ denotes the path integration over all trajectories.


\subsection{Brief review of stochastic thermodynamics}\label{sec:review}

We now quickly review conventional stochastic thermodynamics, in which there is no uncertainty about the
thermodynamic parameters~\cite{van2015ensemble,seifert2012stochastic,rao_esposito_my_book_2019}.
A more extensive review can be found in \cref{app:review_stoch_thermo}.
Readers already comfortable with stochastic thermodynamics can skip to the next section.

Conventional stochastic thermodynamics considers a system with a specified Hamiltonian that is
coupled to $N$ independent infinite thermal reservoirs, which have associated inverse temperatures $\beta^\nu$,
This coupling between the systems and the reservoirs results in the system evolving according to a CTMC, where as described below,
the physical requirement of microreversibility puts constraints on how the rate matrix of the CTMC is related to the physical
parameters of the process. Below we go through the main definitions of relevant quantities used in stochastic thermodynamics. Their
explicit formulas as well as the extension to the case of particle reservoirs are shown in \cref{app:review_stoch_thermo}.

\subsubsection*{Ensemble thermodynamics}
\label{sec:2_ensemble}
The ensemble internal energy is written as \mbox{$U_t := \langle u_t \rangle = \sum_x p_t(x) u_t(x)$}.
The system exchanges energy with each of the reservoirs (e.g., via kinetic molecular collisions). Additionally, one can consider that the system exchanges particles with some of the reservoirs. This extension is discussed in \cref{app:review_stoch_thermo} and used later in one of the examples.
 \emph{The first law of thermodynamics} can be formulated as
\begin{equation}
\Delta U_t = \pmb{Q}_t + \pmb{W}_t
\label{eq:first_law}
\end{equation}
where $\Delta U_t := U_t - U_{t_i}$
$\pmb{Q}_t := \int_{t_i}^t \ud t' \dot{\pmb{Q}}_{t'} = \int_{t_i}^t \ud t' \sum_x \dot{p}_{t'}(x) u_{t'}(x)$ is total heat
flow into the system during the interval $[t_i, t]$,
and $\pmb{W}_t  := \int_{t_i}^t \ud t' \dot{\pmb{W}}_{t'} = \int_{t_i}^t \ud t' \sum_x p_{t'}(x) \dot{u}_{t'}(x)$ is the total work
on the system during that interval.
The heat flow rate $\dot{\pmb{Q}}_t$ can be decomposed into heat flows in
from the separate reservoirs, i.e., direct energy flows in from the separate reservoirs: $\dot{\pmb{Q}}_t = \sum_{\nu}
\dot{\pmb{Q}}^\nu_t$.

The entropy rate can be decomposed as 
\eq{
\Delta S_t = \pmb{\Sigma}_t + \pmb{\mathcal{E}}_t
\label{eq:second_law}
}
where $\Delta S_t = S_t-S_{t_i}$, $\pmb{\Sigma}_t$ is the entropy production (EP) and $\pmb{\mathcal{E}}_t$ is the entropy flow (EF).
%
%
\emph{The second law of thermodynamics} is enforced by the fact that
for any rate matrix, EP rate is non-negative, i.e., $\dot{\pmb{\Sigma}}_t \geq 0$ and
so  ${\pmb{\Sigma}}_t \geq 0$.

When LDB holds, the EF rate can be expressed in terms of \emph{thermodynamic entropy}, i.e.,
\eq{
\dot{\pmb{\mathcal{E}}}_t =  \sum_\nu \beta^\nu \dot{\pmb{Q}}^{\nu}_t\, .
}
In this case, \cref{eq:second_law} means that
\eq{
{\pmb{\Sigma}}_t = \Delta S_t  - \beta {\pmb{Q}}_t
}
if there is only a single heat bath, with inverse temperature $\beta$
By \cref{eq:first_law} this can be written as
\eq{
\dfrac{{\pmb{\Sigma}}_t}{\beta} =  \pmb{W}_t - \left( \Delta U_t - \dfrac{\Delta S_t}{\beta} \right) = \pmb{W}_t - \Delta F_t
\label{eq:7}
}
where $\Delta F_t$ is the difference in the Helmholtz free energy between time $0$ and time $t$.

Suppose  we wish to minimize the work expended over all possible protocols 
$(u_t, K_t)$
acting on such a system, so long as they take some specified initial
pair of an energy function and distribution  $(u^*_{t_0}, p^*_{t_0})$ to some specified ending pair,
$(u^*_{t_f}, p^*_{t_f})$. For that specified pair, $\Delta U_{t_f}$ is fixed, independent of the protocol.
So by \cref{eq:first_law} the minimal such work is $\Delta U_{t_f} - \min_{(u_t, K_t)}  \pmb{Q}^{(u_t, K_t)}$
where $\pmb{Q}^{(u_t, K_t)}$ is the total heat over the interval $[t_i, t_f]$ flowing into the system
if it follows protocol $(u_t, K_t)$. By the second law, that minimum is just $\Delta S_{t_f} / \beta$.

Plugging in, the minimal work over all protocols is $\Delta U_{t_f} - \Delta S_{t_f} / \beta$.
Therefore the quantity on the RHS of \cref{eq:7} is the difference between the actual work expended
to go from $(u^*_{t_0}, p^*_{t_0})$ to $(u^*_{t_f}, p^*_{t_f})$ and the minimal possible.
Accordingly, that quantity is called the \textit{dissipated work}, and by the first law it can be written as
a function of the initial distribution as follows:
\eq{
 \pmb{W}_{\mathrm{diss}}(p_{t_i}) :=  \dfrac{ \Delta S_{t_f}}{\beta}(p_{t_i}) - \pmb{Q}_{t_f}(p_{t_i})
\label{eq:diss_work}
}

In this way \cref{eq:7} connects
a purely thermodynamic quantity to a purely dynamic quantity, defined in terms of rate matrices.
Note that the energy function and distribution are treated specially in the definition of dissipated
work --- no other parameters of the thermodynamic process are allowed to vary when determining
the ``minimal'' work.

Let us now define the \textit{prior} initial distribution as the one that minimizes that function
when \cref{eq:diss_work}. We can consider the directional derivative of
that function, evaluated for any initial distribution, in the direction of any other initial distribution.
In particular, that directional derivative of the dissipated work must equal $0$
when evaluated at the prior initial distribution $q_{t_i}(x)$
in the direction of any other initial distribution, $p_{t_i}(x)$.
This can be used to show that
\eq{
 \!\!\!\!\!\!\!\! \pmb{\Sigma}(p_{t_i}) &=
\pmb{W}_{\mathrm{diss}}(p_{t_i})  \nonumber \\
	&=
\left[D(p_{t_i} || q_{t_i}) - D(p_{t_f} || q_{t_f})\right]\nonumber \\
& \;\;\;+
\pmb{\Sigma}^\alpha(q_{t_i})
\label{eq:mismatch}
}
where the distributions $p_{t_i}(x)$ and $q_{t_i}(x)$ evolve into the distributions
$p_{t_f}(x)$ and $q_{t_f}(x)$, respectively.
$\pmb{\Sigma}(q_{t_i})$ is called the \textit{residual EP}~\footnote{Formally, \cref{eq:mismatch}
assumes that the conditional distribution $P(x(t_f) | x(t_i))$
has a single ``island''; the extension of the analysis here to the case of multiple islands is
straight-forward.
See~\cite{kolchinsky2016dependence,wolpert_thermo_comp_review_2019,wolpert_kolchinsk_first_circuits_published.2020,riechers2021initial}.}.
The thermodynamic process implemented by a given apparatus is thermodynamically reversible iff
the residual EP of the process is zero, and the initial distribution happens to equal the prior of the process.

The drop in KL divergence in \cref{eq:mismatch} is called the
\emph{mismatch cost} of running the process with initial distribution
$p_{t_i}$.
By the data-processing inequality for KL divergence, the mismatch cost is nonzero, and by inspection, it equals zero if the initial distribution equals the prior. So
mismatch cost is the extra EP generated by running the process with initial distribution $p_{t_i}$
rather than the prior $q_{t_i}$,
in addition to the residual EP which would be generated if the process were run with the prior as the initial distribution.

\subsubsection*{Trajectory thermodynamics}
The {\emph{trajectory internal energy}} is written as $\pmb{u}_{t}(\vx)$.
The first law of thermodynamics on the trajectory level for any
time $t$ is
\begin{equation}
\frac{\ud}{\ud t} \pmb{u}_t(\vx) = \dot{\pmb{q}}_t(\vx) + \dot{\pmb{w}}_t(\vx)
\end{equation}
where
$\dot{\pmb{q}}_t(\vx) = \sum_\nu \dot{\pmb{q}}_t^\nu(\vx)$ is the trajectory heat and $\dot{\pmb{w}}_t(\vx) = \sum_x
\delta_{x,\vx(t)} \dot{u}_t{(x)}$ trajectory work.
Trajectory entropy is defined as $s_t(\vx) := - \ln p_t(\vx(t))$. The time derivative of entropy can be decomposed as
\eq{
\frac{\ud}{\ud t} \pmb{s}_t(\vx) &= \dot{\pmb{\sigma}}_t(\vx) + \dot{\pmb{\epsilon}}_t(\vx)
\label{eq:stoch_entr_fixed_rate_matrix}
}
where $\dot{\pmb{\sigma}}_t(\vx)$ is the trajectory EP rate and $\dot{\pmb{\epsilon}}_t(\vx)$ is the trajectory EF rate.
It is straightforward to verify that by averaging these trajectory-level quantities over all trajectories, we recover the
ensemble-level versions, i.e., $\boldlangle {\dot{\pmb{q}}}^\nu_t \boldrangle = \dot{\pmb{Q}}^\nu_t$, $\boldlangle
\dot{\pmb{\epsilon}_t} \boldrangle = \dot{\pmb{\mathcal{E}}}_t$,
and $\boldlangle \dot{\pmb{\sigma}}_t \boldrangle = \dot{\pmb{\Sigma}}_t$.
Due to LDB, trajectory EP can be expressed as
\begin{equation}
\pmb{\sigma}(\vx) = \ln \frac{\pmb{P}(\vx)}{\pmb{P}^\dag(\tvx)}
\end{equation}
where $\pmb{P}(\vx)$ is the probability of observing trajectory $\vx$ and $\pmb{P}^\dag({\tvx})$ is the probability of
observing time-reversed trajectory $\tvx(t) := \vx(t_f-t)$ under time-reversed protocol. Defining $P(\sigma) := \int \mathcal{D} \vx
\vP(\vx) \delta(\sigma - \pmb{\sigma}(\vx))$,
it is straightforward to show that the trajectory EP $\sigma$ fulfills the \emph{detailed fluctuation theorem},
\begin{equation}\label{eq:dft}
\frac{P(\sigma)}{{P}^\dag(-\sigma)} = e^{\sigma}
\end{equation}
where $P^\dag$ denotes the probability under a time-reversed protocol and $\sigma$ is random variable~\cite{van2015ensemble}.
Finally, ensemble EP can be expressed as Kullback-Leibler divergence between probabilities of forward and reversed
trajectories~\cite{Parrondo_2009}
\begin{equation}\label{eq:at}
\pmb{\Sigma} = D_{\vx}(\pmb{P}(\vx)||\pmb{P}^\dag(\tvx))\,.
\end{equation}

\section{Effective thermodynamics}
\label{sec:3}


As described in the introduction, we are interested in how the conventional laws of stochastic thermodynamics concerning
the evolution of a system change when there is uncertainty about the parameters of that evolution, but we
assume LDB holds, whatever those parameters are.

In \cref{sec:effective}, we describe a general framework of effective quantities that are averaged over different apparatuses. 
In \cref{sec:meaning_effective_EP} we continue the description of effective ensemble thermodynamics in the general case where the transition rates
of the systems are not known with infinite precision. Such uncertainty must
arise whenever we do not know the exact number of heat reservoirs, their temperatures/and or chemical potentials.
However, uncertainty
can even arise when we \textit{do} know those quantities exactly, and even when we impose LDB.
This is because even if we knew those quantities to infinite precision, LDB does not uniquely
fix the rate matrix $K$, and so there can still be uncertainty concerning $K$. 

We exemplify the effective thermodynamics approach in \cref{sec:bit_erasure_example} on the example of a bit erasure for the case of fermionic bits. We start with the brief review of system when coupled to a heat bath with certain temperature. Then we continue with the generalized case when the system is coupled to a heat bath with uncertain temperature.


\subsection{Effective quantities}
\label{sec:effective}

Suppose we are given some generic quantity $Y^\alpha$, which depends on
the apparatus $\alpha$, and which
may also depend on time and/or the random trajectory through the system's state space (where any of the dependencies may be
implicit). The probability distribution at time $t$ is denoted as $p_t(x \mid \alpha)$.
Trajectory probability is denoted as  $\vP(\vx \, | \, \alpha)$.

We define the \emph{effective} value of  $Y$ as its expectation over $\alpha$, and write it as
\eq{
\overline{Y} &:= \int \ud P^\alpha \, Y^\alpha
}
where we require that the $\alpha$-dependence of all arguments of $Y^\alpha$ is made explicit.
So for example, if $Y^\alpha$ depends on the initial probability distribution over states, that distribution
occurs in its argument list as $p_{t_0}(x | \alpha)$.
%
%
In particular, the effective state probability at time $t$ is
\eq{
\overline{p}_t(x) &= \int \ud P^{\alpha}\,  p_t(x \mid \alpha)  \\
	&= \int \ud P^{\alpha}\, p_t^\alpha(x)
}
where we use the shorthand $p_t^\alpha(x) := p_t(x|\alpha)$. We also write the effective trajectory probability
up to time $t$ 
\begin{equation}
\overline{\vP}_t(\vx) = \int \ud P^{\alpha} \,  \vP_t(\vx \, | \, \alpha)
\end{equation}

Note that in both scenarios I and II, since the apparatus is fixed throughout $[t_i, t_f]$ once it
is sampled, the joint dynamics over $X \times A$ is given by the master equation,

\eq{
\dot{\overline{p}}_t(x)
	&= \sum_{x'} \int \ud  P^{\alpha'} \, K^{\alpha \alpha'}_{xx'} p_t^{\alpha'}(x')
\label{eq:joint_dynamics}
}

where  $K^{\alpha \alpha'}_{xx'} = K^\alpha_{xx'} \, \delta(\alpha',\alpha)$.
Averaging both sides of \cref{eq:joint_dynamics} over $\alpha$ and interchanging the derivative and average
on the LHS, we get
\eq{
\dot{\overline{p}}_t(x)= \sum_{x'} K_{xx'} \overline{p}_t(x')\,
}
where
\eq{
K_{xx'} := \int \ud  P^{\alpha} \, K^{\alpha \alpha'}_{xx'} p_t(x'|\alpha')
}
So the dynamics over the system considered by itself is simply
the dynamics of a coarse-graining of the joint system-apparatus. Since the transition rate matrix
of that coarse-grained dynamics depends on the probability distribution, the effective dynamics is not described by a linear Markov master equation.

\subsection{Effective ensemble stochastic thermodynamics}
\label{sec:effective_ensemble}

We write  $\pmb{W}^\alpha$ for the ensemble energetic work expended for a specific apparatus $\alpha$
as it sends $p^\alpha_{t_i}$ to  $p^\alpha_{t_f}$. By conservation of energy
\eq{
\pmb{W}^\alpha = \Delta {U}^\alpha -  \pmb{Q}^\alpha
\label{eq:102aa}}
where the change in the internal energy for each apparatus $\alpha$ during the process is
\eq{
\Delta U^\alpha &= \sum_x [p^\alpha_{t_f}(x) u^\alpha_{t_f}(x) - p^\alpha_{t_i}(x) u^\alpha_{t_i}(x)]
\label{eq:23}
}
In addition, by the second law, for any specific apparatus $\alpha$,
\eq{
 \sum_\nu  \beta^{\nu,\alpha} \pmb{Q}^{\nu, \alpha}  \le S(p^\alpha_{t_f}) - S(p^\alpha_{t_i})
}

These quantities are given by integrating over time and fixing a specific apparatus. If we instead average
over apparatuses and fix a specific time, we get the formulas for
the effective ensemble energy and effective ensemble entropy at time $t$:
\eq{
\overline{U}_t &= \int \ud P^\alpha \sum_x p^\alpha_t(x) u_t^\alpha(x) \\
\overline{S}_t &= - \int \ud P^\alpha \sum_x   p^\alpha_t(x) \ln p_t^\alpha(x)
}

If we run over both time and apparatuses, we get the formulas for the effective work on the system and heat transferred to
the reservoir(s) up to time $t$:
\eq{
\overline{\pmb{W}}_t &= \int \ud P^\alpha \int_{t_i}^t \ud t' \sum_x p_{t'}^\alpha(x) \dot{u}_{t'}^\alpha(x) \\
\overline{\pmb{Q}}_t &= \int \ud P^\alpha \int_{t_i}^t \ud t' \sum_x \dot{p}_{t'}^\alpha(x) u_{t'}^\alpha(x)
}
%

In addition, the effective ensemble EF rate is
\eq{
\label{eq:21}
\dot{\overline{\pmb{\mathcal{E}}}}_t &= \int \mathrm{d} P^\alpha \sum_\nu \beta^{\alpha,\nu} \dot{\pmb{Q}}^{\alpha,\nu}_t \\
	&= \int \mathrm{d} P^\alpha \sum_{xx'\nu} K^{\alpha,\nu}_{x,x'} p^\alpha_t(x') \ln
\dfrac{K^{\alpha,\nu}_{x,x'}}{K^{\alpha,\nu}_{x',x}}\,
\label{eq:effective_ensemble_EF}
}
and $\overline{\pmb{\mathcal{E}}}_t =  \int_{t_i}^{t} \ud t \, \dot{\overline{\pmb{\mathcal{E}}}}_t$.
Similarly, the effective ensemble EP is $ \overline{\pmb{\Sigma}}_{t} = \int_{t_i}^{t} dt \dot{\overline{\pmb{\Sigma}}}_t$,
where the effective ensemble EP rate is
\eq{
\label{eq:12a}
\dot{\overline{\pmb{\Sigma}}}_t &= \int \ud P^\alpha \dot{\pmb{\Sigma}}^\alpha_t  \geq 0
}
Plugging in the expression for $ \dot{\pmb{\Sigma}}^\alpha_t $ in terms of rate matrices, we confirm
that
\eq{
\dfrac{\ud \overline{S}}{\ud t} &= \dot{\overline{\pmb{\mathcal{E}}}_t}  +  \dot{\overline{\pmb{\Sigma}}}_t
}
where
\eq{
\dot{\overline{\pmb{\Sigma}}}_t &= \int \mathrm{d} P^\alpha \sum_{xx'\nu} K^{\alpha,\nu}_{x,x'} p^\alpha_t(x')
		\ln \dfrac{K^{\alpha,\nu}_{x,x'}p_t(x')} {K^{\alpha,\nu}_{x',x} p_{t}(x)}
\label{eq:27}
}

Combining these definitions, we can write the first and second law of thermodynamics for effective ensemble quantities
as
\eq{
\label{eq:26}
\Delta \overline{U}_t &= \overline{\pmb{Q}}_t + \overline{\pmb{W}}_t\\
\Delta \overline{S}_t &= \overline{\pmb{\Sigma}}_t + \overline{\pmb{\mathcal{E}}}_t
\label{eq:second_law_effective}
}
respectively, simply by averaging the first and second laws over all $\alpha$.
Note in particular that  since $ \dot{\overline{\pmb{\Sigma}}}_t  \geq 0$, the total effective ensemble EP
generated in the interval is non-negative. This is
why~\cref{eq:second_law_effective} can be identified as the effective scenario's second law.

On the other hand though, \cref{eq:21} shows that in general
the effective ensemble EF rate cannot be expressed in terms of
the expected effective heat flows for the separate reservoirs,
$\dot{\overline{\pmb{Q}^{\nu}_t}}= \int \ud P^\alpha \dot{\pmb{Q}}^{\nu,\alpha}_t$,
despite our assumption that LDB holds for each apparatus separately.
This discrepancy between the two rates would exist even if we knew with certainty that there was only one reservoir, i.e., $N=1$; the discrepancy reflects the fact that there can be statistical
coupling between $\beta^\alpha$ (and $K^{\alpha,\nu}$ and therefore)
$\dot{\pmb{Q}}^\alpha_t$, due to the uncertainty over $\alpha$.
This discrepancy means that
the time-derivative of the effective ensemble entropy is not necessarily lower-bounded by the ensemble
effective heat flow rate.
So the version of the effective scenario's second law of thermodynamics, \cref{eq:second_law_effective} is not as consequential as the standard version
in which there is no uncertainty about $\alpha$.


\subsection{Example: Bit erasure of information stored in fermionic bits with uncertain temperature}
\label{sec:bit_erasure_example}

We now illustrate how environment uncertainty affects the design of an experimenter's protocol as well as the associated EP
for the effective scenario, using the example
of bit erasure of a fermionic bit in finite time  \cite{diana2013finite}. First, we
review the analysis when there is no uncertainty about temperature. Then
we extend that analysis by introducing uncertainty about the temperature
of the single heat bath, and therefore (in order to enforce LDB) in the trajectories of the rate matrices and so
of the energy function.

\subsubsection*{Bit erasure of information stored in fermionic bits}

Our system has two states, labeled $0$ and $1$. The
probabilities of those two states are denoted as $p_t(1) \equiv p(t)$ and $p_t(0) \equiv 1-p(t)$ respectively. In
this subsection, we will denote the dependence of a generic quantity $x$ on time $t$ as $x(t)$ (rather than $x_t$) to clarify the reasoning.
In particular, the chemical potential of an electron in the quantum dot is written as $\mu(t)$.
(Recall that in \cref{app:review_stoch_thermo}
we discuss the extension of stochastic thermodynamics to include chemical potentials.)

For simplicity, we will only consider variations in the trajectory of energy functions across $t \in [t_i, t_f]$, with a fixed
map specifying the rate matrix that goes with each possible energy function.
Specifically, we set the rate matrix at all times $t$ to
\begin{equation}
K^{\beta}(t) =
\left(\begin{array}{cc}
  K^\beta_{00}(t)&K^\beta_{01}(t)\\
  K^\beta_{10}(t)&K^\beta_{11}(t)\\
  \end{array}
\right) =
\left(
\begin{array}{cc}
  -k^\beta(t)&1-k^\beta(t)\\
  k^\beta(t)&-(1-k^\beta(t))\\
  \end{array}
\right)
\label{eq:55}
\end{equation}
 where
 \begin{equation}
k^\beta(t) = \frac{1}{1+\exp(\beta h(t))}
\label{eq:55aaa}
 \end{equation}
and we use the shorthand $h(t)= u(t)-\mu(t)n_1$ for the difference between the energy and
the chemical potential times $n_1 =1$.
Since there is no remaining freedom to vary the trajectory of rate matrices  across $t \in [t_i, t_f]$ once we specify
a trajectory of energy functions across $t \in [t_i, t_f]$, we will sometimes refer to that trajectory of energy functions as
a ``control protocol''.

Given \cref{eq:55}, we can write
\eq{
K^\beta_{10}(t) (1-p^\beta(t)) - K^\beta_{01}(t) p^\beta(t) = k^\beta(t) - p^\beta(t)
\label{eq:56}
}
and so
 \begin{equation}
\dot{p}^\beta(t) = k^\beta(t) - p^\beta(t)
\label{eq:58}
 \end{equation}
The solution is
\begin{equation}
p^\beta(t) = e^{-t}\left(p(t_i) + \int_{t_i}^t \mathrm{d} \tau e^\tau k^\beta(\tau)\right)\, .
\label{eq:57}
\end{equation}

From now on, for simplicity, we take $t_i = 0$. Consider a special type of control protocol chosen so that
\begin{equation}
k^\beta(t) \equiv k(t) = (1-t/t_f)p(0) + t/t_f \delta
\label{eq:59}
\end{equation}
for all $t$, arbitrary $\delta > 0$. Note that the RHS is independent of $\beta$.
Plugging in to \cref{eq:55aaa}, the associated control protocol is
\begin{equation}
h^\beta(t) = \frac{1}{\beta} \ln\left(\frac{1+p(0) t + t\delta}{p(0)(t-1)-t\delta} \right).
\end{equation}
Similarly, plugging in to \cref{eq:57} shows that
\begin{equation}\label{eq:prob}
p(t) = e^{-t}\left(p(0)(2 e^t - t e^t -1) + \delta (1+t e^t-e^t) \right)
\end{equation}
Note that despite the notation, this quantity does not depend on the temperature. Finally, the EP is
\eq{
& \pmb{\Sigma}^\beta \equiv \pmb{\Sigma} = \int_0^t \mathrm{d} \tau \dot{p}(\tau) \left(\ln \frac{1-p(\tau)}{p(\tau)} - \beta
h^\beta(\tau)\right) \nonumber\\
&\; =  \int_0^t \mathrm{d} \tau \left[p(t) (1-k(t)) \ln \left(\frac{p(t) (1-k(t))}{(1-p(t))
k(t)}\right)\right. \nonumber \\
& \qquad\qquad \left.+(1-p(t)) k(t) \ln
   \left(\frac{(1-p(t)) k(t)}{p(t) (1-k(t))}\right)\right]
}
Note that EP also does not depend on temperature, due to the choice of the control protocol $h^\beta(t)$.


\subsubsection*{Bit erasure with temperature uncertainty}
We now consider the variant of this bit erasure scenario where
the experimentalist does not know the temperature exactly but instead has some distribution $\ud P^\beta = p(\beta) \mathrm{d}\beta$.
We
suppose that the experimenter is in full control of the control protocol, but that the same, single protocol will
be used for all apparatuses.

Averaging both sides of \cref{eq:57} gives the formula for the evolution of the effective distribution:
\eq{
\overline{p}(t) = e^{-t}\left(p(t_i) + \int_{t_i}^t \mathrm{d} \tau e^\tau \overline{k}(\tau)\right)\
}
with
 \begin{equation}
\overline{k}(t) = \int \mathrm{d} \beta p(\beta) \,  \frac{1}{1+\exp(\beta h^*(t))}
 \end{equation}
where $h^*(t)$ is the protocol chosen by the experimentalist to be used for all apparatuses. 
Note that this is the same evolution one would get by first averaging both sides of \cref{eq:58},
\begin{equation}
\dot{\overline{p}}(t) = \overline{k}(t) - \overline{p}(t)
\end{equation}
and then solving for $\overline{p}(t)$.
This reflects the fact that in this particular situation, where \cref{eq:56} holds, we can write
$\overline{K(t) p(t)} = \overline{K}(t) \cdot \overline{p}(t)$,
and so the dynamics of the effective distribution is Markovian (in contrast with the general case).
%

Similarly to the no-uncertainty analysis, we assume the transition rate matrix has the form
\begin{equation}
\label{eq:prob2}
\overline{k}(t) = (1-t/t_f)p(0) + t/t_f \delta
\end{equation}
%
%
To this end, we should find a control protocol (independent of $\beta$) such that \cref{eq:prob2} equals \cref{eq:prob}.

Next, define the function 
 \begin{equation}
\Psi(h^*) := \int \mathrm{d} \beta p(\beta)\,  \frac{1}{1+\exp(\beta h^*)}
 \end{equation}
Thus, the control protocol $h^\star(t)$ can be obtained by solving the equation for the transition rate $\bar{k}(t) =
\Psi(h^*(t))$.
By solving the equation, we obtain
\begin{equation}\label{eq:h}
h^*(t) = \Psi^{-1}\left((1-t)p(0) + t \delta\right)
\end{equation}
Note that $h^*(t)$ is not the $\beta$-average of $h^\beta(t)$ which can be expressed as $\bar{h}(t) = h^{\tilde{\beta}}(t)$, where $\tilde{\beta} = \left(\int \mathrm{d} \beta P^\beta 1/\beta\right)^{-1}$ is the harmonic mean of beta (corresponding to the
arithmetic mean of temperature).

In the actual experiment, however, the system is coupled to a bath with the certain temperature
$\beta$ and therefore the transition rate is
\begin{equation}
k^{\beta}_{h^*}(t) = \frac{1}{1+\exp(\beta h^*(t))}
\end{equation}
Thus, the probability distribution of a system coupled to a heat reservoir with temperature $\beta$ using the protocol $h^*$, can be
obtained from the equation
\begin{equation}
\dot{p}_{h^*}^{\beta}(t) = k_{h^*}^\beta(t) - p_{h^*}^{\beta}(t)\, .
\end{equation}

As an illustration, consider the special case where there are two possible temperatures.
The first temperature $T_1 = 1/\beta_1$ occurs with probability $P(T_1)$ and the second temperature $T_2=1/\beta_2$ occurs with probability $P(T_2) = 1-P(T_1)$.
Thus, the distribution $P(T)$ can be expressed as
\begin{equation}
P(T) = P(T_1) \delta_{T T_1} + P(T_2) \delta_{T T_2}\, .
\end{equation}
and therefore
\begin{equation}
\overline{k}(t) =  \frac{P(T_1)}{1+\exp\left(\frac{h^*(t)}{T_1}\right)} +    \frac{P(T_2)}{1+\exp\left(\frac{h^*(t)}{T_2}\right)}
\end{equation}

As usual, the control protocol can be obtained by solving \cref{eq:h}. Moreover,
since that protocol is independent of $T$, by definition the effective ensemble entropy production is
\begin{equation}
\bar{\pmb{\Sigma}}_{h^*} = P(T_1) \, \pmb{\Sigma}^{T_1}_{h^*} + P(T_2) \,  \pmb{\Sigma}^{T_2}_{h^*}
\end{equation}

\begin{figure*}[t]
\begin{center}
\includegraphics[width=0.4\linewidth]{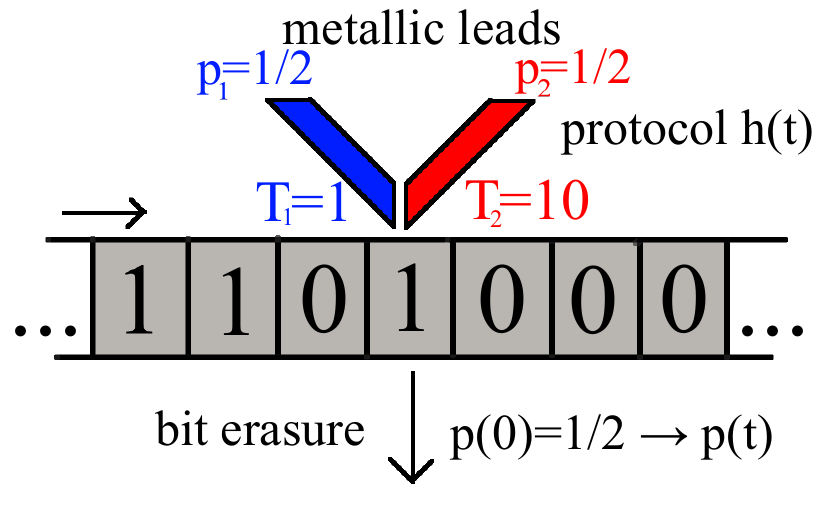}
\includegraphics[width=0.4\linewidth]{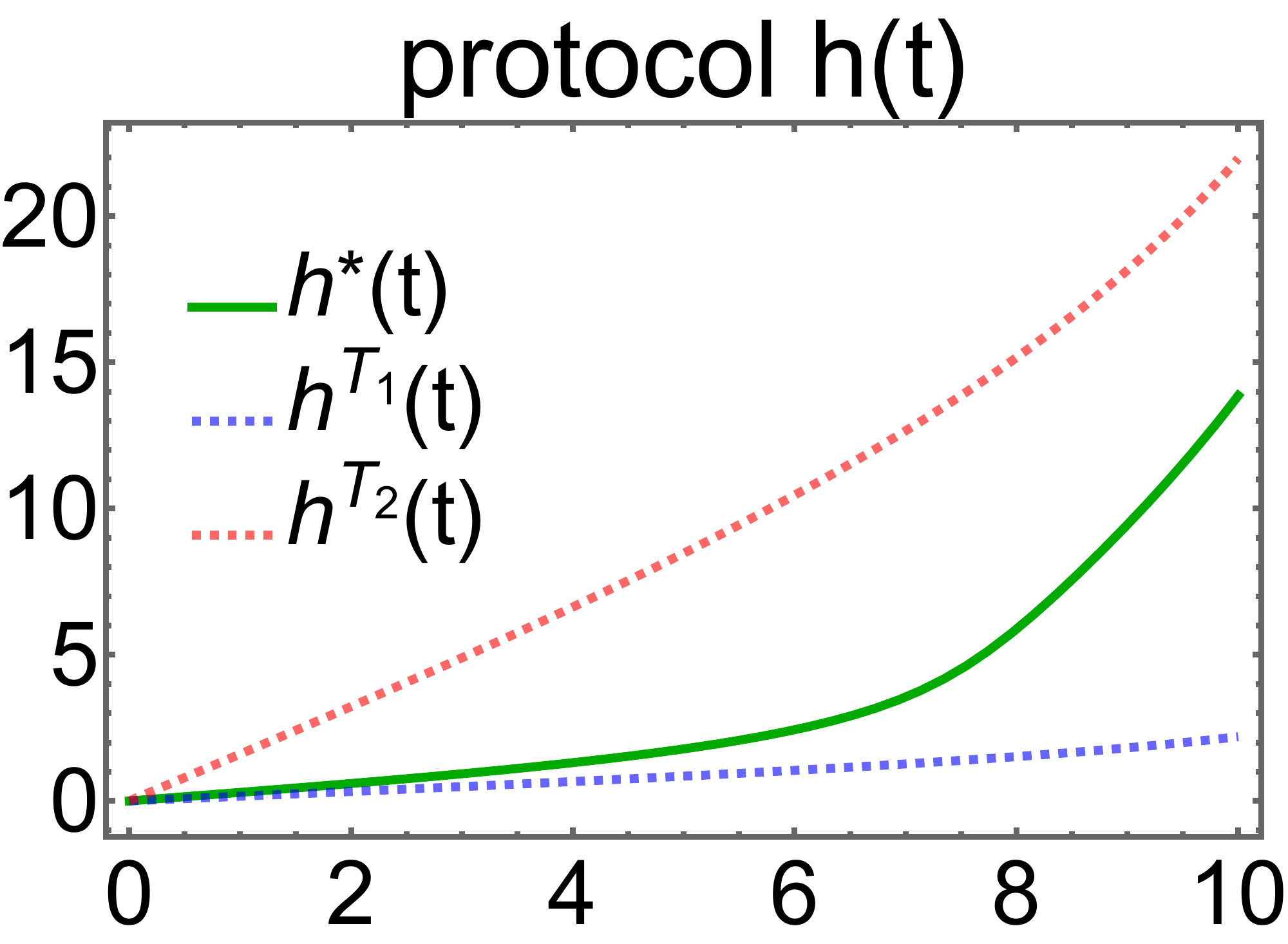}\\
\includegraphics[width=0.4\linewidth]{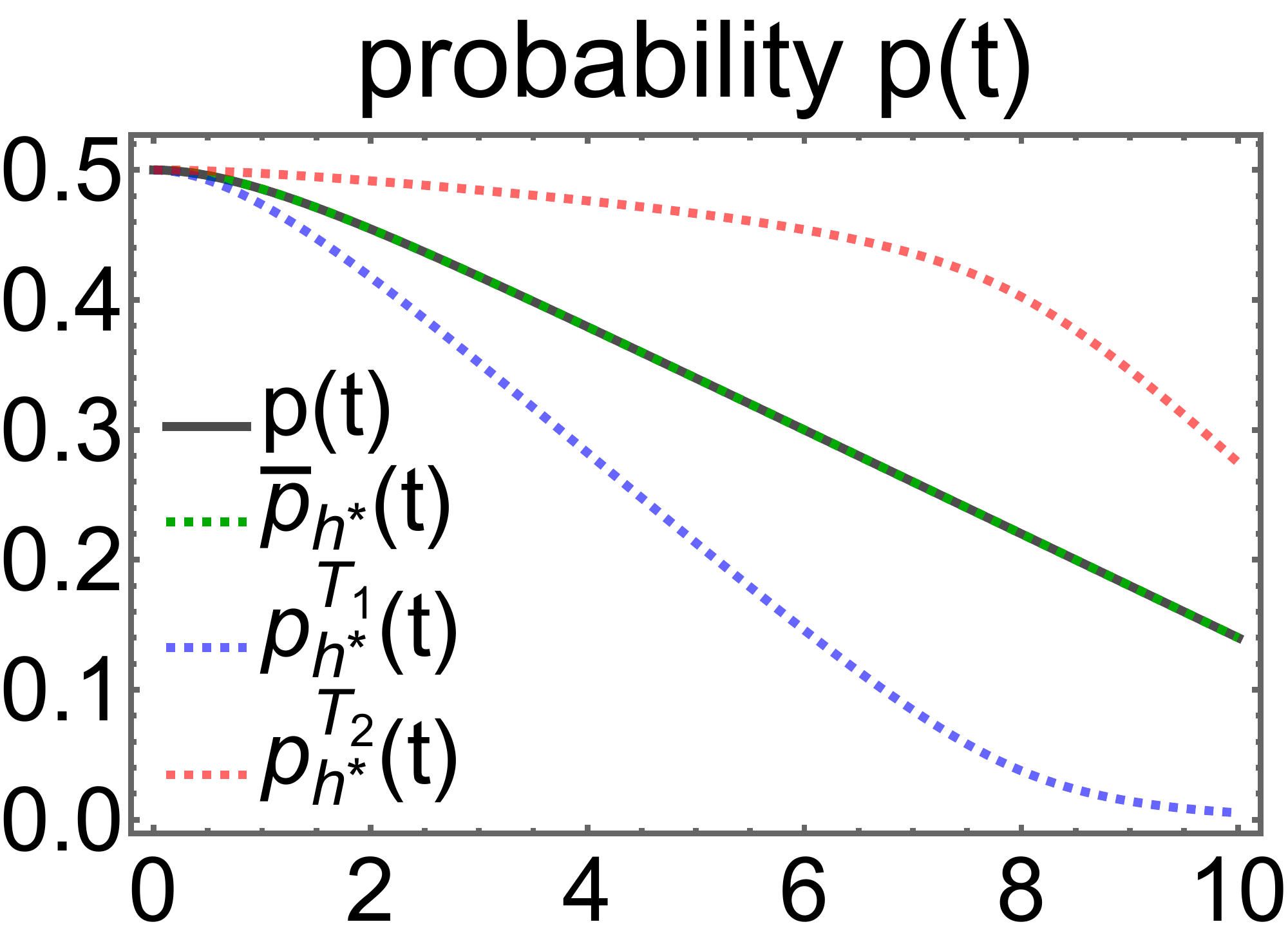}
\includegraphics[width=0.4\linewidth]{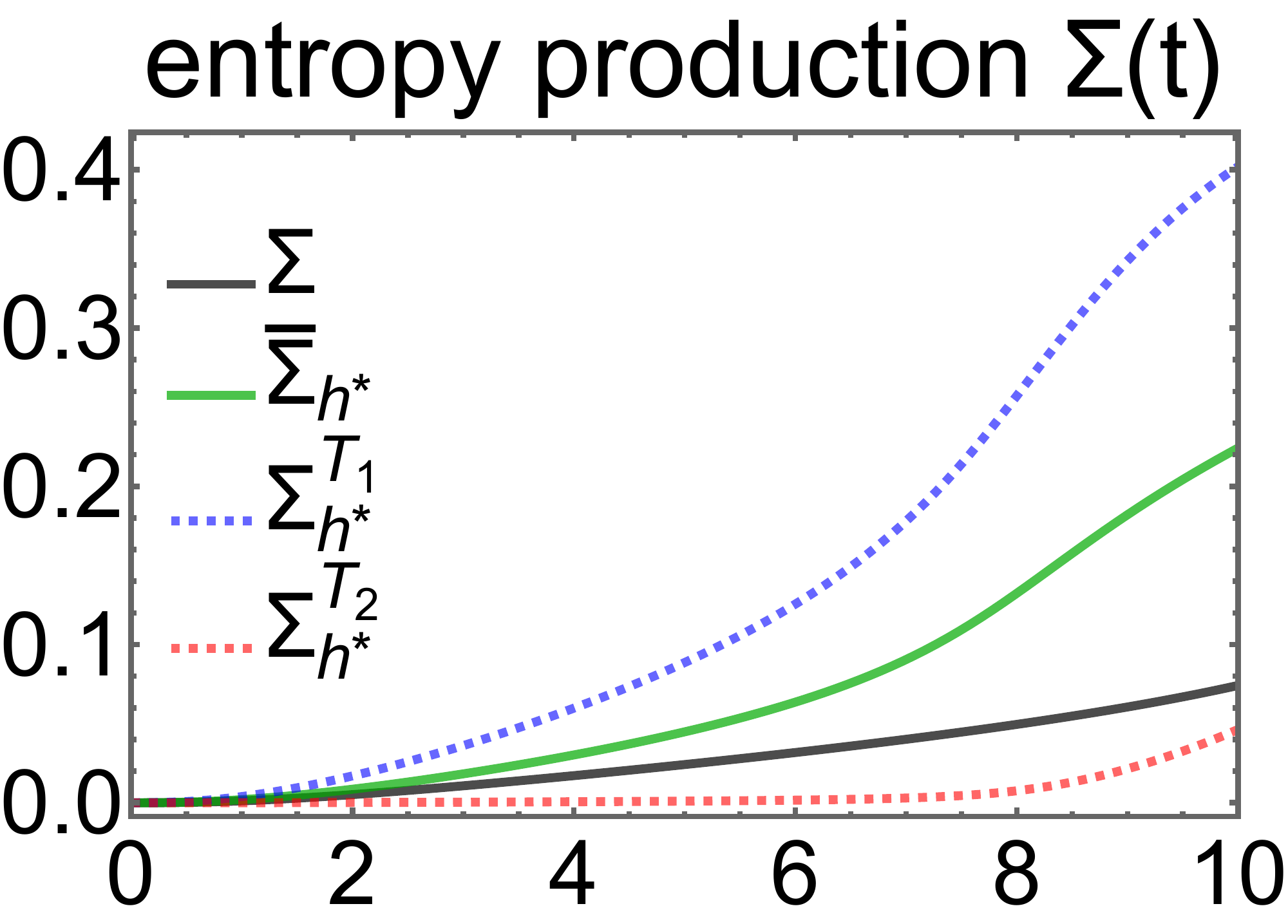}
\caption{Bit erasure of a fermionic bit with uncertain temperature.
The design of the experiment is depicted in the top-left panel. For each site, the bit is coupled to one of the metallic leads with different temperatures $T_1 = 1$ and $T_2=10$ with equal probability. The control protocol for each temperature and for the case of uncertain temperature is depicted in the top-right panel. The actual probability distribution for a given temperature and control protocol is depicted on the bottom-right panel. The entropy production for the case of a certain temperature as well as an uncertain temperature is depicted in the bottom-right panel.
}
\label{fig:3}
\end{center}
\end{figure*}

In \cref{fig:3} we plot the control protocol $h(t)$, probability distribution $p_1(t)$, and entropy production $\pmb{\Sigma}(t)$ for
for this case of bit erasure with uncertainty about which of two possible temperatures the bath has.
We assume parameters $T_1= 1$, $T_2=10$, $t_f = 10$ and $\delta=0.1$, and take
$P(T_1)=P(T_2)=1/2$.

In the top-right panel of \cref{fig:3} we plot $h^*(t)$ along with
the control protocols $h^T(t)$
for the two cases where there is no temperature uncertainty, for the two possible values of $T = \{T_1,T_2\}$.
The expected distribution $\bar{p}(t)$ is displayed in the bottom-left panel, along with
the two distributions $p^T_{h^*}$ which could actually occur in the experiment,
when the experimenter fixes the protocol to $h^*(t)$ and the temperature of the bath is one of $T = \{T_1,T_2\}$.
Finally, we compare the total EP for the case of certain temperature $\pmb{\Sigma}^T(t)$
(which does not  vary with $T$),  with the case of total EP for the case of uncertain temperature given by
$\overline{\pmb{\Sigma}}_{h^*}$
in the bottom-right panel of \cref{fig:3}. Despite the fact that $p(t)$ coincides with $\bar{p}(t)$,
the effective ensemble EP increases to the case of EP for the case of a certain temperature.
For comparison, we depict $\pmb{\Sigma}^T_{h^*}$ for $T = \{T_1,T_2\}$.

The main reason why $\overline{\pmb{\Sigma}}_{h^*}$ is higher than $\pmb{\Sigma}$ is the fact that for the situation when $T_1$ is chosen, the protocol $h^*$ forces the distribution $p_{h^\star}^{T_1}(t)$ to decrease much below $p(t)$, which on one hand makes the bit erasure more efficient (the final probability distribution is closer to $(1,0)$), but one has to pay much more dissipated work and consequently entropy production $\pmb{\Sigma}_{h^*}^{T_1}$ to get the distribution closer to the ideal bit erasure. This work is not compensated in the other case when the temperature is equal to $T_2$. We see that while the entropy production $\pmb{\Sigma}_{h^*}^{T_2}$ is lower than $\pmb{\Sigma}$, this does not compensate enough the entropy production $\pmb{\Sigma}_{h^*}^{T_1}$ and the average entropy production $\overline{\pmb{\Sigma}}_{h^*}$ is larger than $\Sigma$ in the case when the temperature is certain. Let us finally mention that this is a similar situation hidden Markov pump \cite{esposito2015stochastic} when the actual EP (here corresponding to $\pmb{\Sigma}$ calculated from the complete microscopic structure of the hidden pump in lower than when the entropy is calculated for a coarse-grained pump. I.e., the experimenter's knowledge of the setup affects the obtained EP. While in the case of \cite{esposito2015stochastic} the experimenter lacks the precise information about the system, there the experimenter lacks the precise knowledge of the heat bath and its temperature.

\section{The two types of effective dissipated work}
\label{sec:diss_work_types}

In much of the rest of this paper we focus on the case of a single bath. Specifically,
we investigate how the properties of dissipated work 
change from those in  \cref{eq:diss_work} when we that number of baths but introduce uncertainty in the other parameters .

It turns out that there (at least) two natural ways to extend the reasoning that results in \cref{eq:diss_work}
to that case
where the number of baths is fixed to one but there is uncertainty about the
other parameters defining the apparatus. These two extensions reflect two
different ways of defining ``minimal possible effective work''.


\subsection{The adapted and unadapted scenarios}

Write ${\hat{\alpha}_t} := 
(u^\alpha_t, K^\alpha_t)$
to mean the components of $\alpha$ that specify the protocol, evaluated at time $t$.
Similarly write $\underline{\alpha}_t$ to mean all components of $\alpha_t$ \emph{other} than those that specify the protocol,
e.g., the initial probability distribution, the temperature, chemical potentials, and other thermodynamic forces, etc.
Thus, $\alpha_t$ can be decomposed as $\alpha_t = ({\hat{\alpha}_t},\underline{\alpha}_t)$.
Recall though that for simplicity we are assuming that all non-protocol components of $\alpha$ are
time-independent, so we can simplify this to $\alpha_t = ({\hat{\alpha}_t},\underline{\alpha})$.

In the sequel, when we are considering the entire trajectory over times $t$ of the values $(u^\alpha_t, K^\alpha_t)$ 
specified by a particular $\hat{\alpha}$, we write $(\textbf{u}^\alpha, \textbf{K}^\alpha)$, in
keeping with our convention that bold characters indicate trajectories. Similarly,
we write $\hat{\alpha}$ without a subscript to mean the entire trajectory of the protocol, $(\textbf{u}^\alpha, \textbf{K}^\alpha)$.

In the first approach to extending the definition of the minimal possible effective work, we
define it as the least work that could be achieved if the energy component of the protocol, $\textbf{u}$,
were fully fixed in an explicitly known manner by the experimentalist,
\textit{before} the other uncertain parameters, $\underline{\alpha}$, were (randomly) determined.
To make sure that uncertainty in parameters like the temperature do not cause LDB to be violated even though
the parameters other than $\bm{u}^\alpha$ are set independently of $\bm{u}^\alpha$, we must
allow the rate matrix trajectories $\textbf{K}^\alpha$ to be statistically dependent on both
  $\bm{u}^\alpha$ and $\underline{\alpha}$.
So in this approach, we restrict attention to measures over $\alpha$ of the form
\eq{
{\mathrm{d}}P^{\textbf{u}^\alpha} {\mathrm{d}} P^{\underline{\alpha}} {\mathrm{d}} P^{(\textbf{K}^\alpha | \textbf{u}^\alpha, \underline{\alpha})}
\label{eq:58a}
}
with no statistical coupling between the energy component of the
the protocol and the non-protocol parameters defining the apparatus.
(It is the term ${\mathrm{d}} P^{(\textbf{K}^\alpha | \bm{u}^\alpha, \underline{\alpha})}$ in \cref{eq:58a} that allows us to ensure that
LDB holds).

We use the term \textit{unadapted} to refer to measures of this form.
For unadapted measures, the minimal possible work is
\eq{
W^{unad}_{\min} = \min_{{\hat{\alpha}}}  \int \ud P^{\underline{\alpha}}  \big(\Delta U^{{\hat{\alpha}}, {\underline{\alpha}}} -
				 \pmb{Q}^{{\hat{\alpha}}, {\underline{\alpha}}} \big)
\label{eq:unadapted}
}
We also use the term ``{unadapted}'' to refer specifically to
the kind of optimal protocol given in \cref{eq:unadapted}, and to the associated definitions of least possible (effective expected) work
and of dissipated (effective expected) work. In particular, the unadapted dissipated work is the amount of work that could have
been saved if the experimentalist
has intervened to change the marginal distribution $\ud P^{{\hat{\alpha}}}$, leaving all other aspects of the
experimental setting the same. (An example of unadapted dissipated work is presented in \cref{sec:bit_erasure_example}.)
%

In the second approach, we do not compare the actual work to the best that
could have been achieved if the experimentalist had changed $\ud P^{{\hat{\alpha}}}$ to be some
delta function,
leaving $\ud P^{\underline{\alpha}}$ unchanged. Rather
we consider the best possible work that could have occurred ``by luck ``, if there were the best possible
statistical coupling between the protocol and the other thermodynamic parameters, again
leaving $\ud P^{\underline{\alpha}}$ unchanged.
Formally, in this second approach we define the ``least possible'' work for a given $\ud P^\alpha$
to be the minimum of the expected work as one varies over the conditional $P^{\hat{\alpha}|\underline{\alpha}}$.

We refer to this measure over apparatuses as the \textit{adapted}
statistical coupling
between the protocol and the other parameters specifying the apparatus that could have occurred, counterfactually,
for the specified (fixed) marginal over those other parameters.
We also refer to the associated minimal effective ensemble work as the adapted
minimal (effective ensemble) work,
\eq{
W_{\min}^{ad} = \int \ud P^{\underline{\alpha}}  \min_{{\hat{\alpha}}} \big(\Delta U^{{\hat{\alpha}}, {\underline{\alpha}}} -
				 \pmb{Q}^{{\hat{\alpha}}, {\underline{\alpha}}} \big)
\label{eq:leibniz}
}
and refer to the associated dissipated effective ensemble work as the adapted (effective ensemble) dissipated work.

Note that we can self-consistently define the adapted dissipated work for a single apparatus $\alpha$,
\eq{
W^{{\hat{\alpha}}, \underline{\alpha}}_{\mathrm{diss}} &:=
W^{{\hat{\alpha}}, \underline{\alpha}} -  \min_{{\hat{\alpha}'}} \big(\Delta U^{{\hat{\alpha}}', {\underline{\alpha}}} -
				 \pmb{Q}^{{\hat{\alpha}}', {\underline{\alpha}}} \big)
\label{eq:leibniz_single}
}
in the sense that the adapted effective dissipated work is the ${\alpha}$-average of
the adapted dissipated work for a single $\alpha$.
The analogous property does not hold for the unadapted dissipated work in general,
since the min and the integral in \cref{eq:unadapted} will typically not commute. (Indeed, if they did commute,
then the unadapted dissipated work would equal the adapted dissipated work.)

\subsection{Physical interpretation of effective EP for a specific adapted scenario}
\label{sec:meaning_effective_EP}

We now describe a relationship between effective dissipated work and effective
ensemble EP, to provide a thermodynamic interpretation of effective ensemble EP.
In this subsection we focus on
calculating the minimal adapted dissipated work.

For simplicity, we suppose that all the apparatuses $\alpha$ with nonzero probability take
the same initial distribution $p^*_{t_i}(x)$ to
the same ending distribution $p^*_{t_f}(x)$, i.e., for all $\alpha$ with nonzero probability,
 $p^\alpha_{t_i}(x) = p^*_{t_i}(x)$ and $p^\alpha_{t_f}(x) = p^*_{t_f}(x)$.
This means that $\Delta S^{\alpha}$ is independent of $\alpha$.

We similarly assume that
the initial and final energy functions are independent of $\alpha$.
By~\cref{eq:23}, these two assumptions mean that $\Delta U^\alpha$ is independent of $\alpha$, and
so in particular it is independent of the protocol.
So by \cref{eq:leibniz}, the adapted minimal effective ensemble work is
\eq{
\Delta U - \int \ud P^{{\underline{\alpha}}}  \!  \min_{{\hat{\alpha}}}  \pmb{Q}^{{\hat{\alpha}}, {\underline{\alpha}}}
\label{eq:diss_work_2}
}

Also for simplicity, from now on we assume that each reservoir for any specific apparatus $\alpha$ has
the same temperature, so that we can write $\beta^{\nu, \alpha} = \beta^\alpha$.
Under this assumption
the second law reduces to the inequality
\eq{\dfrac{(S(p^\alpha_{t_f}) - S(p^\alpha_{t_i}))}{\beta^\alpha} &=  \dfrac{(S(p_{t_f}) - S(p_{t_i}))}{\beta^\alpha} \\
	& \ge \pmb{Q}^\alpha
}
In general, the minimum of \cref{eq:diss_work_2} will occur when the protocol ${\hat{\alpha}}$ accompanying the rest of the
apparatus $\underline{\alpha}$ saturates the second law. As a result, the adapted minimal work is
\eq{
\Delta U - \int\ud P^{{\underline{\alpha}}} \dfrac{\Delta S^{\underline{\alpha}}} {\beta^{\underline{\alpha}}}
	&= \Delta U - \Delta S \int \ud P^{{\underline{\alpha}}} (\beta^{\underline{\alpha}})^{-1} \\
	&= \Delta U - \Delta S \overline{\beta^{-1}}
\label{eq:adapted_minimal_work}
}
Plugging into \cref{eq:26} and combining,
the adapted dissipated
work is
\eq{
\overline{\pmb{W}}_{\mathrm{diss}}
	&=  {\Delta S} \overline{\left( \beta^{-1}\right)} - \overline{\pmb{Q}}
\label{eq:51a}
}

On the other hand,
multiplying both sides of \cref{eq:leibniz_single} by $\beta^\alpha$ before averaging over $\alpha$,
then again using the fact that $\Delta U^\alpha$ is independent of $\alpha$, we get
\eq{
\overline{\beta \pmb{W}_{\mathrm{diss}}} &= \int \mathrm{d} P^\alpha \, \beta^\alpha[ \pmb{W}^\alpha - \Delta U]
		+ \int \mathrm{d} P^\alpha \beta^\alpha \dfrac{\Delta S}{\beta^\alpha} \\
& = \Delta S  - \overline{\beta \pmb{Q}} = \overline{\pmb{\Sigma}}
\label{eq:18a}
}
Plugging this into \cref{eq:51a} and rearranging gives
\eq{
\overline{\pmb{\Sigma}} =
\left( \dfrac{ \overline{\pmb{W}}_{\mathrm{diss}} + \overline{\pmb{Q}}}{\overline{\left( {\beta^{-1}}\right)}} \right)
		-\overline{\beta \pmb{Q}}
\label{eq:52}
}

So in general, in this setting where all apparatuses map  $p^*_{t_i}(x)$ to $p^*_{t_f}(x)$ and
$\Delta U^\alpha$ is independent of $\alpha$,
the effective ensemble EP is not proportional to
the effective (adapted) dissipated work, in contrast to the no-uncertainty case.
Instead it equals the effective inverse-temperature-weighted dissipated work (\cref{eq:18a}), or alternatively
it is an affine function of the effective dissipated work (\cref{eq:52}).
Note though that if $\beta$ is fixed, independent of $\alpha$, then
the relationship between adapted effective dissipated work and effective EP mirrors their relationship in
the no-uncertainty case:
\eq{
\beta \overline{ \pmb{W}}_{\mathrm{diss}} =  \overline{\pmb{\Sigma}}
\label{eq:44}
}
(Compare to \cref{eq:18a}.)

As a variant of this scenario, suppose instead that we still have a single reservoir, one which does not exchange particles with
the system, and also still suppose that the other
thermodynamic parameters
can vary with $\alpha$. However, now suppose that both $p^{\underline{\alpha}}_{t_i}$ and $p^{\underline{\alpha}}_{t_f}$ can also
vary with $\underline{\alpha}$, unlike before. To have the definition of minimal adapted work still be meaningful in this situation,
assume that the protocol cannot change either of those two distributions, although it can change intermediate distributions.
Formally, this means that the support of
$P(\hat{\alpha} | \underline{\alpha})$ is restricted so that for all $\underline{\alpha}$,
no $\hat{\alpha}$ is possible which affects the initial and final distributions.
(If changing the protocol were allowed to change the initial and / or final distributions, then in general
the minimal adapted work would be arbitrarily negative.)
Under this assumption changes to $\hat{\alpha}$ without any changes to $\underline{\alpha}$
do not change the drop in entropy, $\Delta S^{\hat{\alpha}, \underline{\alpha}}$.
Nor do they change $\Delta U^\alpha$, which is still independent of $\alpha$.
However, now changes to $\underline{\alpha}$ \textit{will} change the drop in entropy.
In this case \cref{eq:51a} gets replaced by
\eq{
\overline{\pmb{W}}_{\mathrm{diss}} &= \overline{\left(\dfrac{\Delta S} {\beta}\right)} - \overline{\pmb{Q}}
}




\subsection{Mismatch cost and effective ensemble EP for a specific unadapted scenario}
\label{sec:mismatch_cost_effective}

A common setting in which we have uncertainty about thermodynamic parameters is where
we know everything about the apparatus with complete certainty  --- the number of reservoirs
and their parameters, the paths followed by the Hamiltonian and the rate matrix, etc. --- except
that we do not know the {initial distribution} that is run
with those parameters. Concretely, this setting arises whenever
we have a fixed physical apparatus that is run with a randomly generated initial distribution. (So we
are concerned with unadapted dissipated work.)
One common example is a computer that is used by different users; each user
implicitly fixes a distribution over the inputs to the computer, i.e., fixes its the initial
distribution over the states of the computer. Other common examples are a
single cell floating in different environments, or a fixed digital gate
that can be positioned at different locations in a digital circuit (and so have different
distributions over the inputs it receives from the rest of the circuit).
%
%
For simplicity, for the rest of this subsection take  $\beta =1$.


In the situation under consideration, where we are certain about the temperature of the process, effective dissipated work
and effective EP are identical, given by \cref{eq:44}. By applying the formula for the mismatch cost \cref{eq:mismatch}, we get the formula for the effective mismatch cost
\eq{
\overline{\pmb{W}}_{\mathrm{diss}} &= \overline{\pmb{\Sigma}}  \nonumber \\
 &\!\!\!\!\!\!\!\!\!\!\! = \int \ud P^\alpha \bigg[D(p^\alpha_{t_i} || q^\alpha_{t_i}) - D(p^\alpha_{t_f} || q^\alpha_{t_f})
+ {\Sigma}^\alpha(q^\alpha_{t_i}) \bigg]
\label{eq:expected_mismatch}
}
In particular, the expected effective mismatch cost is
\eq{
\overline{\pmb{W}}_{\mathrm{diss}} &= \int \ud P^\alpha \left[D(p^\alpha_{t_i} || q^\alpha_{t_i}) -
				D(p^\alpha_{t_f} || q^\alpha_{t_f})\right] \nonumber \\
	& = \int \ud P^\alpha \left[D(p^\alpha_{t_i} || q^\alpha_{t_i}) - D(R^\alpha p^\alpha_{t_i} || R^\alpha q^\alpha_{t_i})\right]
}
where $R^\alpha$ is the transition matrix of the apparatus $\alpha$, with entries $P^\alpha(x(t_f) | x(t_i))$.

This effective mismatch cost is strictly positive in general since the integrand is always non-negative and only
equals zero in degenerate cases.
In particular, in the situation considered in this subsection, the rate matrix trajectory is
independent of $\alpha$, and therefore so is the prior. So the
effective mismatch cost reduces to
\eq{
\overline{\pmb{W}}_{\mathrm{diss}}
	&= \int \ud P^\alpha \left[D(p^\alpha_{t_i} || q_{t_i}) - D(R p^\alpha_{t_i} || R q_{t_i})\right]
\label{eq:49}
}

As an illustration, suppose that $R$
takes all initial distributions to the same ending distribution (i.e., $R p^\alpha_{t_i} = R q_{t_i}$).
For example, this happens in bit erasure, or in complete
relaxation of a system to its stationary state.
In this case the ending KL divergence is zero, no matter what $p^\alpha_{t_i}$ is.
Since the KL divergence is a convex function of its arguments, this means that
the prior that would minimize effective mismatch cost
is just the effective initial distribution, $q_{t_i}(x) = \overline{p}_{t_i}(x)$.

If we plug this into \cref{eq:49} we can
evaluate the minimal value of the contribution to EP arising from uncertainty
about the initial distribution. Since
effective residual EP is non-negative, we see that
\eq{
\overline{\pmb{\Sigma}}  \ge \int \ud P^\alpha D(p^\alpha_{t_i} || \overline{p}_{t_i}) = D_{JS}(\{p^\alpha_{t_i}\}, P^\alpha)
\label{eq:lower_bound}
}
where $D_{JS}(\{p^\alpha_{t_i}\}, P^\alpha)$ denotes the Jensen-Shannon divergence among the set of distributions
$\{p^\alpha_{t_i}\}$ distributed according to $P^\alpha$. So minimal
EP is strictly positive, as long as $P^\alpha$ puts nonzero probability
mass on at least one distribution $p^\alpha \ne \overline{p}$.
This provides a strengthened version of the second law, applicable whenever there is
uncertainty about the initial distribution, and all initial
distributions with nonzero probability get mapped to the same final distribution.



\section{Control protocols in a specific unadapted scenario}
\label{sec:4}

As discussed at the end of \cref{sec:2_ensemble}, along with \cref{sec:diss_work_types,sec:meaning_effective_EP},
in no-uncertainty stochastic thermodynamics, the ``dissipated work'' is defined as
the difference between two amounts of work. The first is the actual work under a given protocol
that maps an initial distribution $p_{t_i}(x)$ to a final distribution $p_{t_f}(x)$. The second
is the least possible amount of work that would be required to implement that map under
any (counterfactual) protocol, with all thermodynamic parameters other than the protocol,
left the same. One way to modify this definition of dissipated work for the case of uncertain thermodynamic parameters
was investigated at the beginning of \cref{sec:meaning_effective_EP}. In that investigation, the initial and
final distributions were both fixed, independent of the
(non-protocol components of the) apparatus $\underline{\alpha}$. We also
supposed that
 $P(\hat{\alpha} | \underline{\alpha}) = P({\bm{u}, \bm{K}} | \alpha^{-(\bm{u}, \bm{K})})$ was optimized for extracting work,
whether due to conscious intervention by the experimentalist or just by chance.
That optimizing conditional distribution was
called the adapted protocol.

In the next subsection we introduce a modification of that scenario considered in \cref{sec:meaning_effective_EP}.
In the following subsection, we analyze the dissipated work for that modified scenario.

\subsection{Modifying the adapted scenario}

First we list the modifications, and then discuss their formal subtleties:
%
\begin{enumerate}
\item One of the reasons for interest in dissipated work in no-uncertainty stochastic thermodynamics
is because often the experimentalist can intervene in their experiment,
in an (essentially) arbitrary way, even if they do not know $\underline{\alpha}$.
In the language of \cref{sec:diss_work_types}, this is an unadapted scenario. Formally,
in such a scenario
the distribution over the set of counterfactual protocols,  $P(\hat{\alpha} | \underline{\alpha})$,
is independent of $\underline{\alpha}$,
since the experimentalist cannot choose the protocol they implement
to match the other uncertain parameters specifying the apparatus (by definition of their being uncertain about those
other parameters). Physically, this means that
%
the experimentalist can set the protocol --- but can only do so before the other thermodynamic parameters are generated by
sampling the distribution over apparatuses, in a way that is independent of the protocol they have set.
\label{item:II:1}


\item
\label{item:II:2}
Often the experimentalist does not in fact have complete freedom to vary the entire protocol arbitrarily. Often they
will be able to set the trajectory of the energy function directly, but cannot directly set the trajectory of
rate matrices. Instead, for each different apparatus, LDB will constrain the relationship between
the rate matrix and the quantity that the experimentalist \textit{can} control, the energy function.

\item
\label{item:II:3}
Often,  the experimentalist will want to only consider the counterfactual situations that result in some pre-specified final distribution, $p^*(x)$.
However, they will be uncertain about
the initial distribution, as well as other thermodynamic parameters that govern the dynamics, like the temperature.

\end{enumerate}

\cref{item:II:1} means that the protocol of the apparatus is statistically independent of
the other thermodynamic parameters:
\eq{
P(\alpha) = P({\hat{\alpha}})  P(\underline{\alpha})
}

To ensure \cref{item:II:2} while requiring that LDB holds for each apparatus we make several
simplifying assumptions. First, we
assume that there is only a single reservoir in all of the
apparatuses that have a nonzero probability of occurring and that the reservoir
only exchanges energy with the system, not particles. So
the only uncertainty parameter concerning the reservoirs that is relevant to ensuring
LDB is the single number $\beta^\alpha$ (which we will sometimes write as
$\underline{\alpha}^\beta$). In addition,
we suppose that there is a single-valued function $M: (u_t, \beta^\alpha) \rightarrow K^\alpha(t)$, a function which
the experimentalist can set arbitrarily (subject to the constraint of LDB), and so will know with zero uncertainty.
%
So by ``varying the set of counterfactual protocols'' we
vary the marginal distribution $P(\hat{\alpha}^{\bm{u}})$ and the function $M$,
with
\eq{
&P\left( K^{\alpha}(t),\underline{\alpha} | {u^\alpha_t} \right) \nonumber \\
	&\qquad= \delta\left(K^\alpha(t)  - M({u^\alpha_t}, \underline{\alpha}^\beta))\right) P(\underline{\alpha})
}
In our analysis below, it will also be convenient to assume that the support of
$\ud P^{\underline{\alpha}^\beta}$ has a finite minimum (so that there is a maximal
possible temperature), and to require
that $M$ be differentiable with bounded derivative.

There are various formal challenges that arise in the analysis, depending on
the precise definition of \cref{item:II:3}.
Here we clarify \cref{item:II:3} to mean that \textit{averaged over apparatuses}, the final distribution is $p^\ast$, rather
than requiring it have that form for all $\underline{\alpha}$ with nonzero probability. That then leaves the formal challenge
of ensuring that we can meet this
\cref{item:II:3}. If the protocol is chosen so that at least one of the apparatuses is
non-infinitesimally off-equilibrium at $t_f$, then if that apparatus happens to occur,
in general the state distribution $p_t(x)$ would pass through
$p^*(x)$ transitionally at the time $t_f$, after which $p_t(x)$ would keep evolving to other distributions,
no matter what the protocol happens to be then. Indeed,
in general, it may be that two apparatuses $\underline{\alpha}, \underline{\alpha'}$ both cause the state distribution to pass through $p^*(x)$ at $t_f$, but
result in different state distributions at all other times, both before and after $t_f$. This is a complication that
doesn't occur in the conventional, no-uncertainty scenario for the case of a single reservoir, where the counterfactual protocol is completely arbitrary, and so the optimal
counterfactual protocol results in the system being at thermal equilibrium at all times.

To see how to address this problem with ensuring that the condition in \cref{item:II:3} is met, recall the conventional, two-step
 \{quench; semi-static protocol\} process often considered in analyses of the minimal thermodynamic
cost, in which there is no uncertainty, and one needs to set a protocol to change $(u_{t_i}, p_{t_i}) \mapsto (u_{t_f},
p_{t_f})$~\cite{parrondo2015thermodynamics,hasegawa2010generalization}.
Here we restrict attention to protocols that expand that conventional two-step process,
into a three-step \{quench; equilibrate; semi-static protocol\} process. In other words, we insert an intermediate step
into the conventional process, a step in which
we wait long enough for the system to relax to equilibrium, no matter what its temperature, before
starting the step with the semi-static protocol.
(We are assured of being able to do this due to our restrictions on the support of $\ud P^{\beta^\alpha}$.)
Due to our assumption
that the function $M(.)$ has a bounded derivative, we can choose the semi-static protocol in which $u(x)$ varies
in that subsequent third step slowly enough
so that (the rate matrix evolves slowly enough so that) the system is always at equilibrium throughout that step,
again no matter what $\alpha$ is.

As a formal point, we also assume that the very first energy function, $u_{t_i}$, cannot be chosen by the
experimentalist, i.e., it is randomly distributed as specified by $P(\alpha)$. However, the experimentalist
``takes over'' the specification of the trajectory of the energy function starting with
the energy function that the system is quenched to, the one immediately following $u_{t_i}$.

As in the conventional no-uncertainty version of the two-step protocol, we require that the distribution over
states does not change during the quench step of this expanded, three-step
protocol. This is true no matter what the temperature of the
bath is. This means both that there is no heat exchange with the bath in that step, and
that there is no change in Shannon entropy in that step,
no matter what $\alpha$ is. So this step is thermodynamically reversible; if we
inverse-quenched
right away, we would return to the initial distribution and energy $u$, with zero net work expenditure.
Moreover, since the system is at equilibrium throughout the semi-static evolution step,
that step is also thermodynamically reversible; we could run it backward, returning to
the energy and (equilibrium) distribution at the beginning of that step,
and there would be zero net work. This is also true no matter what $\alpha$ is.

On the other hand, in contrast to the quench and semi-static steps, in general, the intermediate, equilibration step
will result in dissipated work for each $\alpha$ considered by itself. More precisely,
while no work is extracted during that step, in a counterfactual process work could have
been extracted, if the protocol could be tailored for that $\alpha$.
%
%

The first of the following subsections contain a preliminary analysis of
this dissipated work. In the subsection after that, we investigate some related issues that arise when a feedback control protocol
is used to define a ``thermodynamic value of information about the actual apparatus''.

\subsection{Optimal work extraction when the temperature is uncertain}
\label{sec:optimal_work_extraction}


Write the Boltzmann distribution for arbitrary energy function $u$ and inverse temperature $\beta^\alpha$ as
\eq{
\pi^\alpha_u(x) = \dfrac{e^{-\beta^\alpha u(x)}}{Z(u, \beta^\alpha)}
}
Also write $u^\alpha_{t_i}$ for the initial, pre-quench energy function for the case where the apparatus is $\alpha$,
and write $u_q$ for the energy function that the experimentalist chooses for the system to quench to.
Similarly, write $u_t$ (with no $\alpha$ superscript) for the energy function at times $t$
during the semi-static evolution (times at which the
experimentalist chooses the energy function).

With this notation the protocol of the process for each $\alpha$
(which is determined before the process starts) can be written as the following:

\begin{center}
\begin{tikzpicture}[node distance = 1.4cm, thick]%
  \node (1) {$(p^\alpha_{t_i},u^\alpha_{t_i})$};
  \node (2) [below=of 1] {$(p^\alpha_{t_i},u_q)$};
  \node (3) [right=of 2] {$(\pi^\alpha_{u_q},u_q)$};
  \node (4) [right=of 1] {$(p^\alpha_{t_f},u_{t_f})$};
  \draw[->] (1) -- node [midway,left] {quench} (2);
  \draw[->] (2) -- node [midway,below] {equilibrate} (3);
  \draw[->] (3) -- node [midway,right] {semi-static ev.} (4);
\end{tikzpicture}
\end{center}
where due to the semi-static nature of the evolution in the third leg, $p^\alpha_{t_f} = \pi^\alpha_{u_{t_f}}$.
As a shorthand, it will also be convenient to write the non-equilibrium internal energy and free energy for arbitrary
energy $u$, state distribution $p$, and apparatus $\alpha$ as
\eq{
U(p,u) &:= \sum_{x} p(x) u(x) \\
F^\alpha(p,u) &:=  U(p,u)  - \dfrac{S(p)}{\beta^\alpha}
\label{eq:90}
}
where in general, $u$ and / or $p$ may depend on $\alpha$.

Note that
our requirement that  $\int \ud P^\alpha  p^\alpha_{t_f}(x) = p^*(x)$ means that
\eq{
 p^*(x) &= \int \ud P^\alpha \pi^\alpha_{u_{t_f}}(x)  \\
	&= \int \ud P^\alpha  \dfrac{e^{-\left[\beta^{\alpha} u_{t_f}(x)\right]}} {Z(u_{t_f}, \beta^\alpha)}
}
This imposes a strong condition on the ending energy function that the experimentalist chooses.
Indeed, in the no-uncertainty case, this requirement would uniquely fix that energy function, up to an overall additive constant.
%
%

Since no work is expended or extracted during the equilibration process $(p_{t_i},u_q) \mapsto (\pi^\alpha,u_q)$,
the total work is given by summing the work during the quench and semi-static evolution steps. Whatever
the protocol is, for each $\alpha$ this total work is
\eq{
\pmb{W}^\alpha &= \underbrace{U(p^\alpha_{t_i},u_q) - U(p^\alpha_{t_i},u^\alpha_{t_i})}_{\mathrm{quench}} \nonumber \\
	&\qquad\qquad+ \underbrace{F^\alpha(p^\alpha_{t_f},u_{t_f}) -
F^\alpha(\pi^\alpha_{u_q},u_q)}_{\mathrm{semi-static \ ev.}}
\label{eq:totw}
}
Adding and subtracting
$S(p^\alpha_{t_i})/\beta^\alpha$ to \cref{eq:totw} and rearranging, we can rewrite this total work as
\eq{
\pmb{W}^\alpha = &F^\alpha(p^\alpha_{t_f}, u_{t_f}) - F^\alpha(p^\alpha_{t_i}, u_{t_i}) \nonumber \\
	   &+F^\alpha(p^\alpha_{t_i}, u_q) - F^\alpha(\pi^\alpha_{u_q}, u_q)
\label{eq:96}
}
The expected work is the $\alpha$-average of this expression.

The first term on the RHS of \cref{eq:96}, the difference between the nonequilibrium free
energies immediately preceding the quench and at the end of the process, is independent of $u_q$.
So the experimentalist cannot affect this term by appropriate choice of protocol.
The second term is instead the change in nonequilibrium free energy that arises in
going from just before to immediately following the equilibration step.

To evaluate that second term, first, note that
since no work is done on the system in the equilibration step,
the change of the expected energy is due to the heat flow, i.e., $\Delta U = \pmb{Q}$. So
the expected value of that second term is
\eq{
\label{eq:96-}
& \int \ud P^\alpha \left[F^\alpha(p^\alpha_{t_i}, u_q) - F^\alpha(\pi^\alpha_{u_q}, u_q) \right]   \\
& \qquad = \int \ud P^\alpha \left[\frac{1}{\beta^\alpha}\left(S(\pi^\alpha_{u_q} - S(p^\alpha_{t_i})\right) -   {Q}^\alpha\right] \\
&\qquad =
\overline{\pmb{W}}_{diss}
\label{eq:96a}
}
where abusing notation, ${Q}^\alpha$ is defined to be the heat flow occurring in just the second, equilibration step,
for apparatus $\alpha$.
\cref{eq:96a} establishes that the ($\alpha$-average of the) second term in  \cref{eq:96}
is just the effective dissipated work of the equilibration step, and therefore the effective dissipated work of the entire process.
(See \cref{sec:meaning_effective_EP}.)


Because $\pi^\alpha_{u_q}$ is the Boltzmann distribution for energy function $u_q$ and inverse temperature $\beta^\alpha$,
as usual in equilibrium thermodynamics,
\eq{
F^\alpha(\pi^\alpha_{u_q}, u_q) = -\dfrac{\ln Z(u_q, \beta^\alpha)}{\beta^\alpha}
}
In addition, $S(p^\alpha_{t_i})$
is independent of $u_q$. Combining, the $u_q$
that minimizes the expression in \cref{eq:96-} is the one that minimizes
$$\int  \ud P_{\alpha}  \,  \left[ \sum_x p^\alpha_{t_i}(x)  u_q(x) + \frac{1}{\beta^\alpha} \ln Z(u_q, \beta^\alpha)\right]$$

To solve for the $u_q$ minimizing this expression, take $ \frac{\partial}{\partial u_q(x)}$ for
each $x$ and set it to zero. This gives a set of coupled equations that $u_q(x)$ must satisfy:
\eq{
\label{eq:3}
\int  \ud P_{{\alpha}}  \,  \left[p^\alpha_{t_i}(x) -  \dfrac{e^{-{{\beta^\alpha}} u_q(x)}}
{\sum_{x'} e^{-{{\beta^\alpha}} u_q(x')}} \right] = 0
}
i.e.,
\eq{
\label{eq:3-}
\overline{p}_{t_i}(x) &= \int  \ud P_{{\alpha}}  \, \dfrac{e^{-{{\beta^\alpha}} u_q(x)}} {\sum_{x'} e^{-{{\beta^\alpha}} u_q(x')}}  \\
	&= \int \ud P^\alpha \pi^\alpha_{u_q}(x)
}
where
\eq{
\overline{p}_{t_i}(x) := \int  \ud P_{{\alpha}}  \,  p^\alpha_{t_i}(x)
}
We denote this optimizing solution to \cref{eq:3-} as $u^*_q$. Note that it is independent of $u_{t_i}, u_{t_f}$ and $p^*$.

As an example, suppose there is no uncertainty in the temperature, only in the initial distribution. Then we
could immediately invert \cref{eq:3-}
to get
\eq{
\label{eq:82-}
u^*_q(x) &=  -\ln \overline{p}_{t_i}(x) / \beta
}
up to an irrelevant additive constant.
Note that this is also the energy function that would be optimal
if with zero uncertainty we knew that the initial distribution were
$\overline{p}_{t_i}(x)$.

More generally, suppose that there is both a nonzero minimal value and a finite maximal value of $\supp (\ud
P_{\beta^\alpha})$,
i.e., a nonzero minimal temperature and a finite maximal temperature that have nonzero probabilities under $\ud P_\alpha$.
Then up to an overall additive constant, there is one and only one solution $u_q$ to \cref{eq:3}
for any given $\overline{p}_{t_i}$ and $\ud P_\alpha$.
(This is proven in \cref{app:proof_of_unique_H*}.)

By plugging $u^*_q(x)$ into \cref{eq:96} and averaging over $\alpha$, we
get a formula for the expected work
during the full protocol:
\eq{
\overline{\pmb{W}}=  &  \int \ud P^\alpha \left[F^\alpha(p^\alpha_{t_f}, u_{t_f}) -
	F^\alpha(p^\alpha_{t_i}, u_{t_i})\right] \nonumber \\
	&+\int \ud P^\alpha \left[F^\alpha(p^\alpha_{t_i}, u_q^*) -
				F^\alpha(\pi^\alpha_{u_q^*}, u_q^*) \right]
\label{eq:102}
}
The first line in \cref{eq:102} is the expected work that would be required if
we were able to use a different (optimal) protocol that was set by the realized value of $\alpha$, i.e.,
it is the minimal work, discussed in \cref{sec:meaning_effective_EP}.
So the second line in \cref{eq:102} is the extra work required due to our \textit{not} being able to
use such a protocol that depends on the realized value of $\alpha$. In other words, it is the extra
work due to our having limited information about the random apparatus.
This kind of increase in the minimal work if the protocol is prevented from depending on the realized value of a random
variable is sometimes called ``thermodynamic value of information'' in the
literature~\cite{parrondo2015thermodynamics,sagawa2013role,sagawa2009minimal,kolchinsky2020entropy,kolchinsky2018semantic}.
What differs between our analysis here and the
previous analyses is that we are concerned with the thermodynamic value of information of
the parameter $\alpha$, whereas those previous analyses instead considered the thermodynamic  value
of information on the initial state of the system.

\subsection{Dynamics of the thermodynamic value of information}
\label{sec:second_minimal_EP_section}

In this subsection, we modify the scenario considered in \cref{sec:optimal_work_extraction}
in two ways. First, we assume that the experimenter can only set the protocol starting at a time $\tau > t_i$,
whereas  $\ud P_\alpha$ is sampled at $t_i$. (We refer to the protocol during the interval starting at $\tau$
as the ``experimenter's protocol''.) Second, we assume that
we know with certainty that $\beta^\alpha = 1$ for all $\alpha$ during that time  of
the experimenter's protocol, following $\tau$. In particular,  $\beta^\alpha = 1$
during the equilibration process. So the only uncertainty the experimenter faces
in choosing their protocol is in what the distribution is when that protocol takes over the dynamics.

In general, the system will evolve between $t_i$ and $\tau$,
potentially according to an uncertain rate matrix.
Our concern is with how the size of the gap $\tau - t_i$ affects the minimal work
that must be expended during the
experimenter's protocol, in light of that evolution between $t_i$ and $\tau$. In particular, we investigate how the
thermodynamic value of knowing the precise $\alpha$ in that interval depends on the length of that interval, i.e., the
derivative with respect to $\tau$ of that value of information.

Since there is no uncertainty in the temperature during the experimenter's protocol,
we can plug into \cref{eq:82-} with $t_i$ replaced by $\tau$,
to see that the optimal energy function for the experimenter to quench to at time $\tau$ is
\eq{
u_q^*(x) &= -\ln \overline{p}_{\tau}(x)
\\
	&=  -\ln \int \ud P^\alpha p_\tau^\alpha(x)
\label{eq:100}
}
up to an overall additive constant~\footnote{Note that even though the temperature during the
experimenter's protocol has no uncertainty, that is not the case during $[t_i, \tau)$.
As a result, ${p}^\alpha_\tau$ can have values different from $\pi_{u^*_q}$,
depending on $P^\alpha$. In fact, even if temperature were fixed during $[t_i, \tau)$,
uncertainty in the initial distribution ${p}^\alpha_{t_i}$  would again mean that
 ${p}^\alpha_\tau$ can have values different from $\pi_{u_q}$.}.
%

If the experimenter had had no uncertainty when they perform the quench, the protocol
starting at $\tau$ would have resulted in zero dissipated work.
Accordingly, we define the \emph{thermodynamic value of information} for this scenario as the
effective dissipated work during the experimenter's protocol that does arise, since there in fact
\textit{is} uncertainty.

Recall that by the analysis in \cref{sec:optimal_work_extraction},
this dissipated work arises during the equilibration that starts immediately
following the time of the quench, which in our current scenario is the time $\tau$.
This dissipated work is given by modifying \cref{eq:96-} to reflect the fact that there
is no temperature uncertainty during the experimenter's protocol:
\eq{
\int \ud P^\alpha \left[F^\alpha(p^\alpha_{\tau}, u^*_{q}) - F(\pi_{u_{q}^*}, u^*_q) \right]
\label{eq:92}
}
where $u^*_q$ is given by the solution to
 \cref{eq:100}.
Expanding the two free energies and
plugging in, we can write this effective dissipated work as
\eq{
I_{{\tau}} &= S\left(\overline{p}_{\tau}\right) - \overline{S}_\tau  \\
	&=   S\left (\int \mathrm{d} P^\alpha p^\alpha_\tau\right) - \int \mathrm{d} P^\alpha S(p^\alpha_\tau)
\label{eq:113}
}
(See the discussion at the end of \cref{sec:optimal_work_extraction}.)

This is the Jensen-Shannon divergence of the set of distributions $\{p^\alpha(x)\}$, weighted according to $\ud P_\alpha$.
Furthermore, since $\alpha$
is a random variable, we can re-express each distribution $p^\alpha(x)$ as a conditional distribution $p(x | \alpha)$.
This allows us to also identify $I_{{\tau}}$ as the mutual information at time $\tau$ between $x$ and $\alpha$,
mutual information that vanishes by the end of the equilibration step.
So $I_{{\tau}}$ is the usual expression for the thermodynamic value of information found in previous analyses in the
literature~\cite{parrondo2015thermodynamics,sagawa2009minimal,wolpert_arxiv_beyond_bit_erasure_2015}.
Here though it reflects the thermodynamic value in knowing what the initial distribution is, while in those previous analyses, it
reflects the value in knowing what the precise initial state is~\footnote{As an aside, in this paper we are considering the
thermodynamic value of information in scenarios in which the (expected) target distribution,
$p_{t_f}(x)$ is fixed, independent of the initial distribution. An alternative is to consider scenarios in which the
conditional distribution, $p(x(t_f) | x(t_i))$ is fixed, independent of the initial distribution. Such scenarios include,
for example, all computational systems more complicated than simple
bit erasure. (In the case of computational systems, uncertainty about the initial distribution amounts to uncertainty about
which of a set of possible users of the computer will set its initial state in any given run of that
computer.) In general, to implement a fixed conditional distribution $p(x(t_f) | x(t_i))$ requires
there to be ``hidden states'', in addition to $X$
Nonetheless, here too the Jensen-Shannon divergence characterizes the value of information in knowing
the actual initial distribution~\cite{wolpert_arxiv_beyond_bit_erasure_2015}, just like in \cref{eq:113}.}.

We wish to evaluate the derivative of $I_{\tau}$ with respect to $\tau$.
To begin, we consider the case where the rate matrix $K_{xx'}(t)$ during the interval $t \in [t_i, \tau)$ has
no uncertainty. We can see how the thermodynamic value of information of $\alpha$ depends on $\tau$ in this case,
by taking the derivative of the RHS of \cref{eq:113} with respect to $\tau$. In \cref{app:value_of_i} we show that the value of
information in this case, $I_{\tau}$, is monotonically decreasing in time, i.e.,
\begin{equation}
\frac{\ud I_\tau}{\ud \tau} \leq 0
\end{equation}
stopping to fall only when
the system reaches equilibrium. (We can derive
the same result from the data-processing inequality.) Physically, this means that the thermodynamic value
of information of $\alpha$ decreases the longer we wait to use it.

In addition to the $\tau$-dependence of the thermodynamic value of information of $\alpha$,
one might also be interested in how the total effective EP (i.e., effective dissipated work),
generated during the entire interval $[t_i, t_f]$, changes with
changes to $\tau \in [t_i, t_f]$. There are two
contributions to that EP: the thermodynamic value of information $I_\tau$, generated during the equilibration
step, and the EP generated earlier, during the interval $[t_i, \tau)$. So to get the derivative of the total effective EP
generated during the entire interval $[t_i, t_f]$, we must add two terms. The first is the derivative of the effective EP under $K(t)$,
i.e., the  effective EP rate, evaluated at
$t = \tau$. The second is the time derivative of the value of information evaluated at that time.

From \cref{eq:12a}, that effective EP rate at $\tau$ is
\eq{
\dot{\overline{\pmb{\Sigma}}}_\tau = \int \ud P^\alpha \,  \sum_{x,x'} K_{xx'}(\tau) p_\tau^\alpha(x') \ln \dfrac{K_{x'x}(\tau)
p_\tau^\alpha(x')}{K_{x'x}(\tau)
p^\alpha_\tau(x)}
}
Adding this to \cref{eq:104}
of \cref{app:value_of_i} gives the derivative of the total effective EP
generated during $[t_i, t_f]$ as
\eq{
\dfrac{d \overline{\Sigma}} {dt} =
\sum_{x,x'} K_{xx'}(\tau) \overline{p}_\tau(x') \ln \dfrac{K_{xx'}(\tau)
			\overline{p}_\tau(x')} {K_{x'x}(\tau) \overline{p}_\tau(x )}
}
This is the EP rate generated by the effective dynamics of $\overline{p}$, evaluated at $\tau$.
So the derivative of the effective EP generated during $[t_i, t_f]$ with respect to the time $\tau \in [t_i, t_f]$
at which the experimentalist takes over the protocol is the phenomenological EP generated at $\tau$.

As a final comment, suppose that $\alpha$ indexes not just the distribution, but also the rate matrix, i.e., suppose that we are unsure about the
dynamics as well as the distribution. This means that
\eq{
\dfrac{d \overline{p}_\tau(x)}{d\tau} &= \int \ud P^\alpha \dfrac{d {p}^\alpha_\tau(x)}{d\tau} \nonumber\\
	&= \int \ud P^\alpha \sum_{x'} K^\alpha_{xx'}(\tau) {p}^\alpha_\tau(x')
}

In this case, the value of information of $\alpha$ can \textit{increase} with $\tau$.
As a simple example, suppose that the initial distribution is independent of $\alpha$,
i.e., $p^\alpha_{t_i}(x) = p_{t_i}(x)$. Then $I_{p_{t_i}} = 0$. Suppose as well that $\alpha$ has two possible values,
as does $x$, and that $dP_\alpha$ is uniform over $\alpha$'s two values.
Finally, also suppose that for $\alpha = 0$, the associated trajectory of rate matrices $K^0_{xx'}(t)$ sends $p_{t_i}(t)$ to the
ending distribution $p^0_\tau(x) = \delta(x, 0)$ with arbitrarily high accuracy, while for
$\alpha = 1$, the associated trajectory of rate matrices $K_{xx'}^1(t)$ sends $p_{t_i}(x)$ to the
ending distribution $p^1_\tau(x) = \delta(x, 1)$ with arbitrarily high accuracy. Then $I_{p_{\tau}}$
is arbitrarily close to $\ln 2$, which establishes the claim.

To illustrate the richness of value of information due to uncertainty, we also consider the value of information for the case where there is uncertain temperature, and therefore (imposing LDB) uncertain rate matrices leading up to the time that the experimentalist's protocol takes over. It is unknown what the general necessary and sufficient conditions
would be for the value of information due to uncertainty to be
nondecreasing as $\tau$ increases. This is a potentially fruitful
area of future research. To exemplify this fact, we calculate the value of information for the example from \cref{sec:bit_erasure_example}, i.e., the bit erasure for the case of the uncertain temperature. We observe that the value of information increases with time, until it reaches its maximal value around $t \approx 8$ and then starts to decrease.

\begin{figure}
\includegraphics[width=0.8\linewidth]{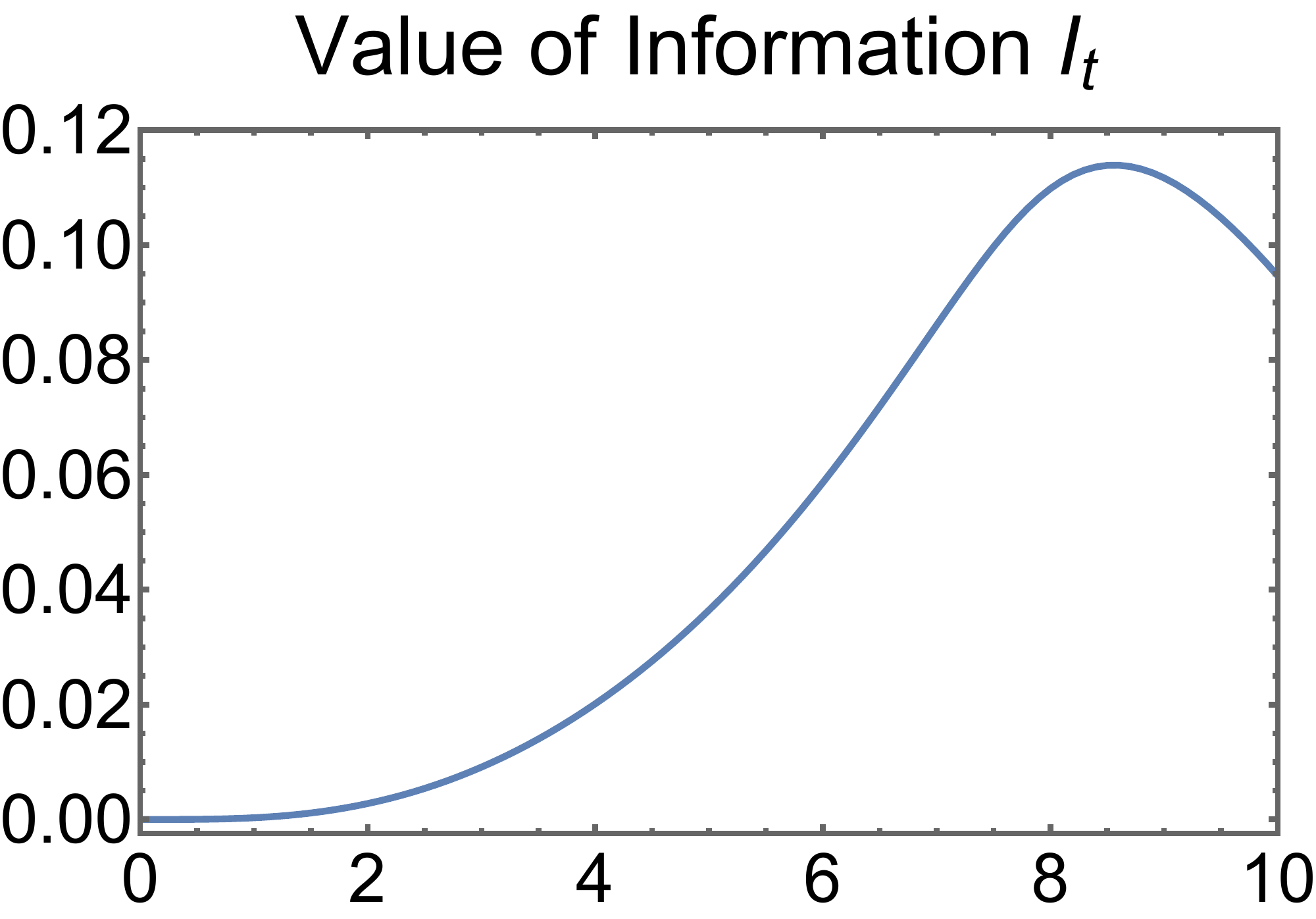}
\caption{Value of information for the case of bit erasure with uncertain temperature from \cref{sec:bit_erasure_example}.}
\label{fig:voi}
\end{figure}

\section{Phenomenological EP and fluctuation theorems}
\label{sec:5}

We now focus on the second, phenomenological scenario, where the
apparatus changes after each stochastic trajectory is generated. In this scenario we are not able
to measure $\vP(\vx|\alpha)$ for any (unknown) apparatus $\alpha$, but only the average probabilities $\overline{\vP}(\vx)$.

\subsection{EP in the phenomenological scenario}

To begin, recall that the $\alpha$-average of the effective EP can be expressed as
\eq{
\overline{\pmb{\Sigma}} = \int \ud P^\alpha D(\vP(\vx \, | \, \alpha)||\ (\vP^\alpha)^\dag(\tvx | \, \alpha))\, .
}
Writing $\ud P^\alpha = p(\alpha) \ud \alpha$ as shorthand,
the joint trajectory probability is $\vP(\vx,\alpha) = \vP(\vx|\alpha) p(\alpha)$.
So
the $\alpha$-averaged effective ensemble EP
is
\eq{
{\overline{\pmb{\Sigma}}} &= D_{\vx,\alpha}(\vP(\vx, \alpha)||\tvP(\tvx, \underline{\alpha}))\nonumber\\
&= \int \ud \alpha \sum_{\vx} \vP(\vx, \alpha) \ln \frac{\vP(\vx, \alpha)}{\tvP(\tvx, \underline{\alpha})}
\label{eq:kl}
}

Now by Bayes' theorem,
\eq{
P(\alpha|\vx) = \frac{\vP(\vx | \alpha)}{\ovP(\vx)} p(\alpha) =  \frac{\vP(\vx, \alpha)}{\ovP(\vx)}
}
where we used the fact that $\ovP(\vx) = \int \ud P^\alpha \vP(\vx | \alpha) $.
Combining this with the chain rule for KL divergence~\cite{gomez2008lower}, we derive
\begin{equation}
\bar{\pmb{\Sigma}} = D_{\vx}(\overline{\vP}(\vx)||\overline{\vP}^\dag(\tvx)) + D_{\vx,\alpha}(\vP(\alpha|\vx)||\tvP(\alpha|\tvx))
\label{eq:45a}
\end{equation}
For later use, introduce shorthand for the first term on the RHS of~\cref{eq:45a},
\eq{
\pmb{\Phi} :=  D_{\vx}(\overline{\vP}(\vx)||\overline{\vP}^\dag(\tvx))
\label{eq:45aa}
}
which we call \emph{phenomenological EP}.
Phenomenological EP measures the irreversibility of
the dynamics quantified by the effective (i.e., apparatus-averaged) probabilities.

Similarly, define the second term on the RHS of~\cref{eq:45a} as
\eq{
\pmb{\Lambda} := D_{\vx,\alpha}(\vP(\alpha|\vx)||\tvP(\alpha|\tvx))
\label{eq:45aaa}
}
which measures the difference between distributions of $\alpha$ estimated from forward and reverse trajectories.
Like many forms of EP, $\pmb{\Lambda}$ is a Kullback-Leibler divergence and is zero only if the forward probability is the same as the backward probability. For these reasons, we refer to $\pmb{\Lambda}$ as \emph{likelihood EP}, even though it need not have a straightforward relation to
dissipated work.

All three EPs in \cref{eq:45a}, \cref{eq:45aa}, and \cref{eq:45aaa} have associated trajectory-level versions:

\eq{
\pmb{\sigma}(\vx | \alpha) &:= \ln \frac{\vP(\vx|\alpha)}{\tvP(\tvx|\alpha)}\\
\pmb{\phi}(\vx) &:= \ln \frac{\overline{\vP}(\vx)}{\overline{\vP}^\dag(\tvx)}\\
\pmb{\lambda}(\alpha|\vx) &:= \ln \frac{{\vP}(\alpha|\vx)}{{\tvP}(\alpha|\tvx)}
\label{eq:last}
}
The first EP corresponds to the effective scenario, where we randomly fix the apparatus and generate an infinite
set of trajectories for that apparatus. The second EP corresponds to the phenomenological scenario, where the apparatus randomly changes
after generating each trajectory.

In general, there will both be $\alpha$ for which $\pmb{\sigma}(\vx | \alpha) > \pmb{\phi}(\vx)$
and $\alpha$ for which  $\pmb{\sigma}(\vx | \alpha) < \pmb{\phi}(\vx)$. However,
\cref{eq:45a} means that
\begin{equation}
\overline{\pmb{\Sigma}} \geq \pmb{\Phi}\, .
\end{equation}
So fixing $\alpha$ to the same value all stochastically generated
trajectories, calculating the associated trajectory-averaged EP, and then averaging
over the unknown values
of $\alpha$ increases the ensemble EP, compared to the case where we average over
apparatuses to calculate EP, and only then average over apparatuses.

The last of these trajectory-level EPs in~\cref{eq:last}, the likelihood EP, is the difference between the first two.
While the likelihood EP is a log-likelihood ratio, in contrast to the common case in which log-likelihood ratios are based on
the same data but different parametric models,
this one is based on the same set of apparatuses, but on forward, resp. reversed trajectories.

The trajectory version of \cref{eq:45a} can be written as
\begin{equation}
\pmb{\sigma}(\alpha,\vx) = \pmb{\phi}(\vx) + \pmb{\lambda}(\alpha|\vx)\, .
\end{equation}
For the first EP $\sigma$, we get to derive the ordinary fluctuation theorem for a given
$\alpha$. For the case of phenomenological EP and likelihood EP, we define the following probabilities:
\eq{
P(\phi) &:= \int \mathcal{D} \vx \, \overline{\vP}(\vx)  \delta(\phi- \pmb{\phi}(\vx))\\
P(\lambda_{\vx}) &:= \int \ud P^\alpha \, \delta(\lambda_{\vx} - \pmb{\lambda}(\alpha|\vx))
}
we can write down detailed fluctuation theorems:
\eq{
\frac{P(\phi)}{{P}^\dag(-\phi)} &= e^\phi\, ,\\
\frac{P(\lambda_{\vx})}{{P}^\dag(-\lambda_{\tvx})} &= e^{\lambda_{\vx}}\, .
}
The consequence of the detailed fluctuation theorem is the integrated fluctuation theorem, i.e, $\langle e^{-\phi} \rangle = 1$,
resp. $\langle e^{-\lambda_{\vx}} \rangle =1$. The first of those IFTs implies that $\langle \phi \rangle \equiv \Phi \geq 0$.
The second of those IFTs means that
\eq{
\pmb{\Lambda}_\vx = \langle \lambda_{\vx} \rangle = \int \ud P(\lambda_{\vx}) \lambda_{\vx} \geq 0
}
We call this \emph{the second law of inference}. It tells us that for each trajectory $\vx$, the $\alpha$-averaged log-likelihood
$\int \ud P^\alpha \ln \vP(\alpha|\vx)$ is larger than the $\alpha$-averaged likelihood obtained from the time-reversed trajectory.
the reverse process.
Of course, by averaging over all trajectories with probability
$\overline{\vP}(\vx)$, we obtain  $\pmb{\Lambda} = \boldlangle \pmb{\Lambda}_\vx \boldrangle \geq 0$.

\subsection{Example: two-state system with uncertain temperature}
\begin{figure*}
\includegraphics[width=0.32\linewidth]{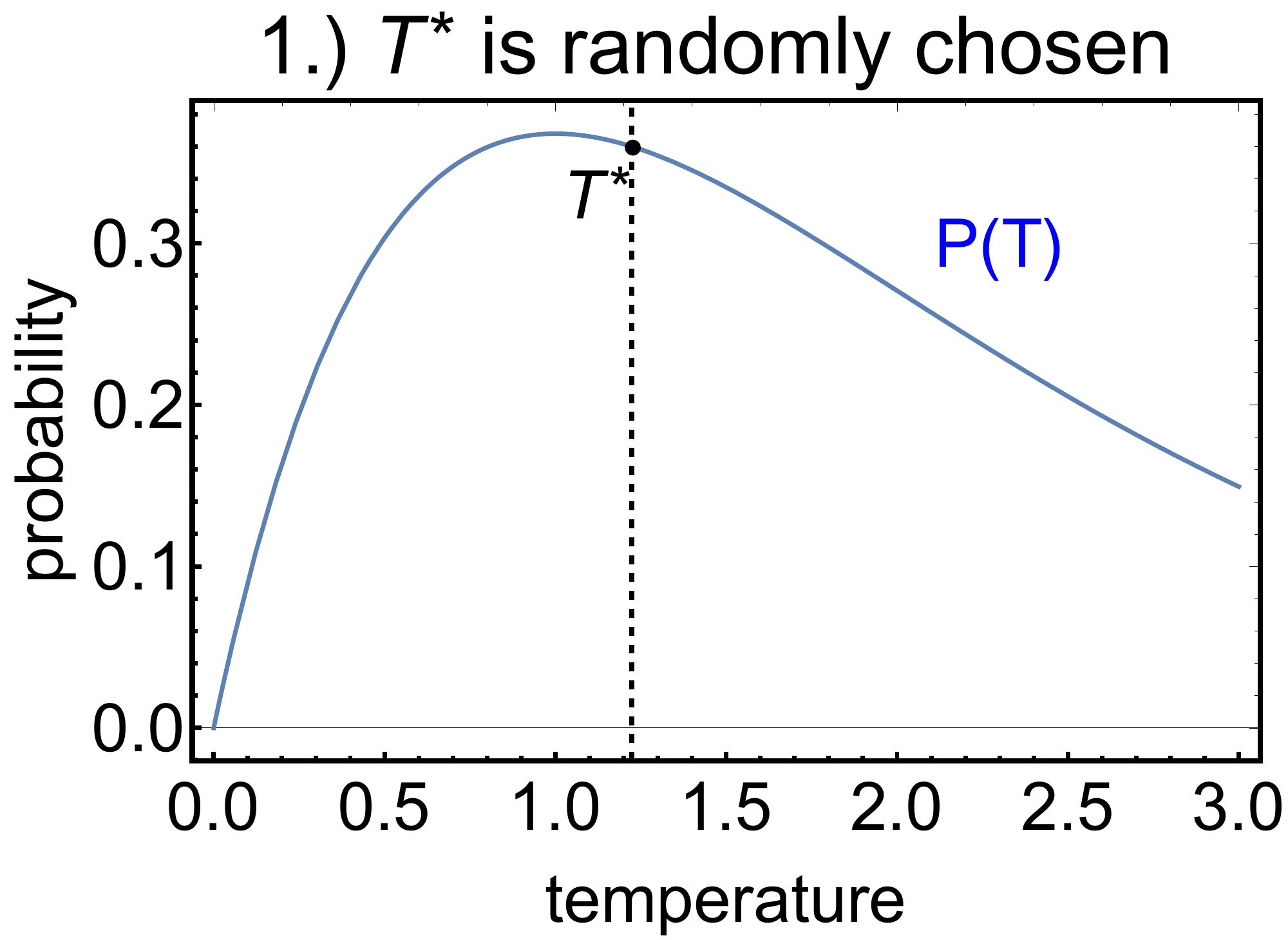}
\includegraphics[width=0.32\linewidth]{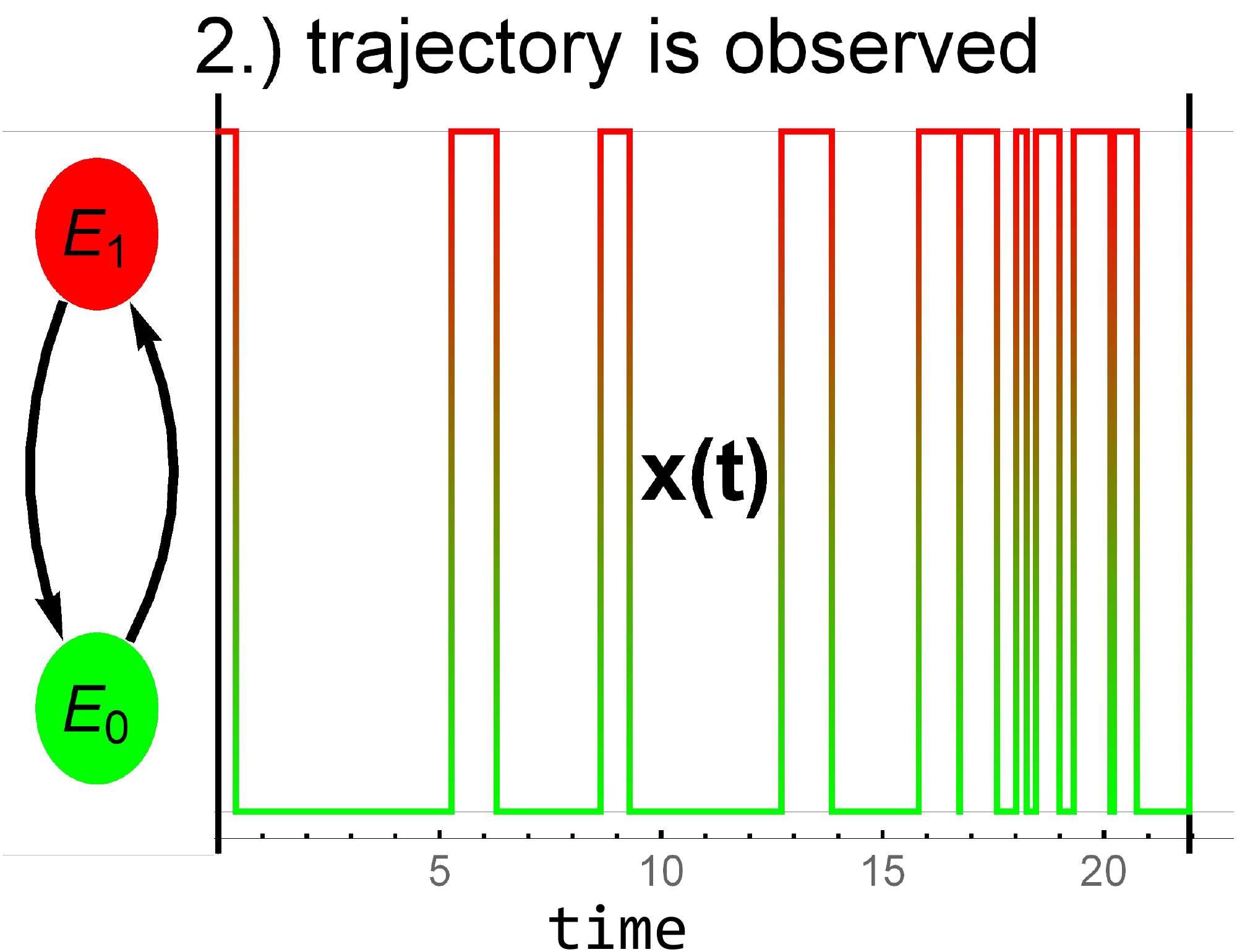}\\
\vspace{5mm}
\includegraphics[width=0.32\linewidth]{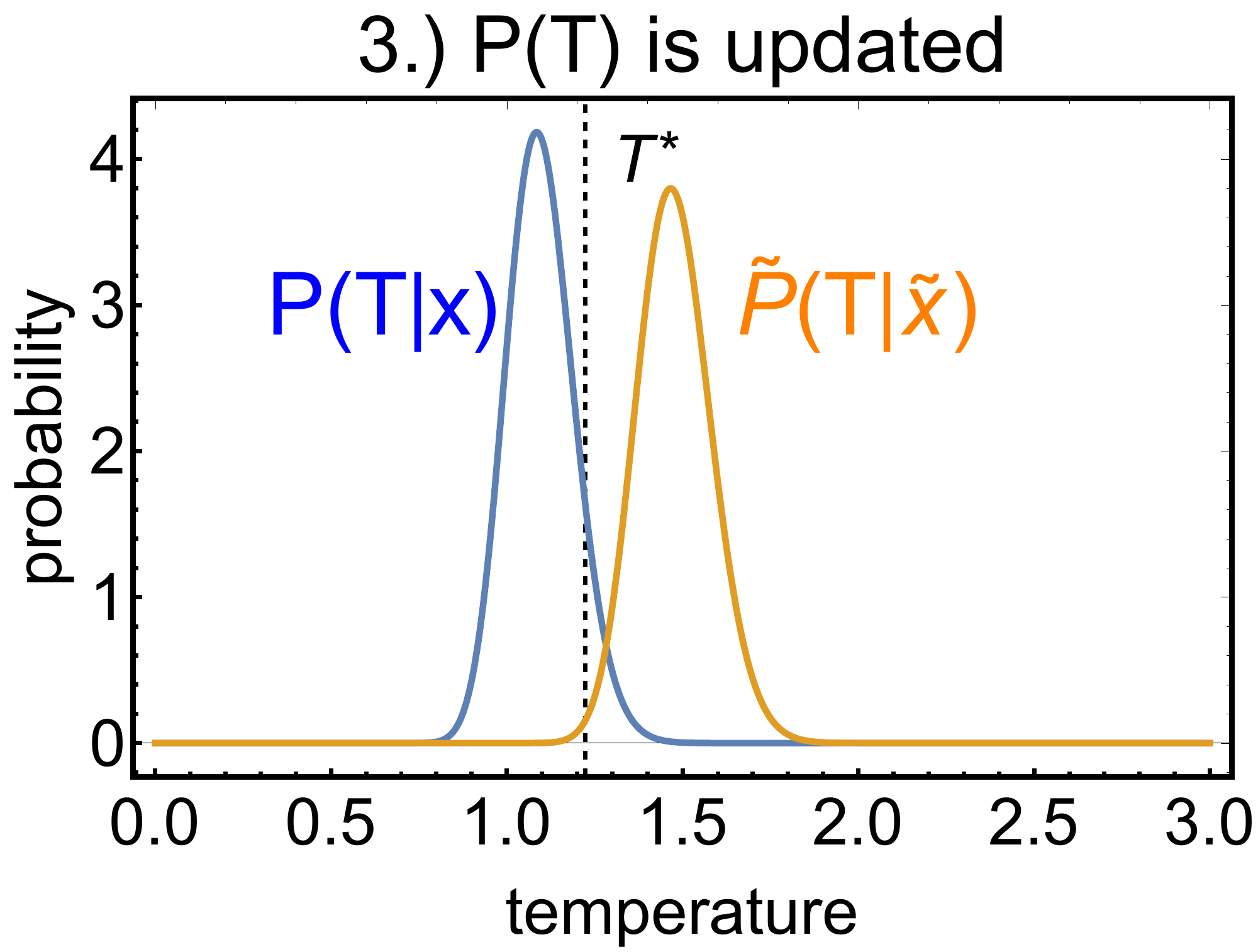}
\includegraphics[width=0.32\linewidth]{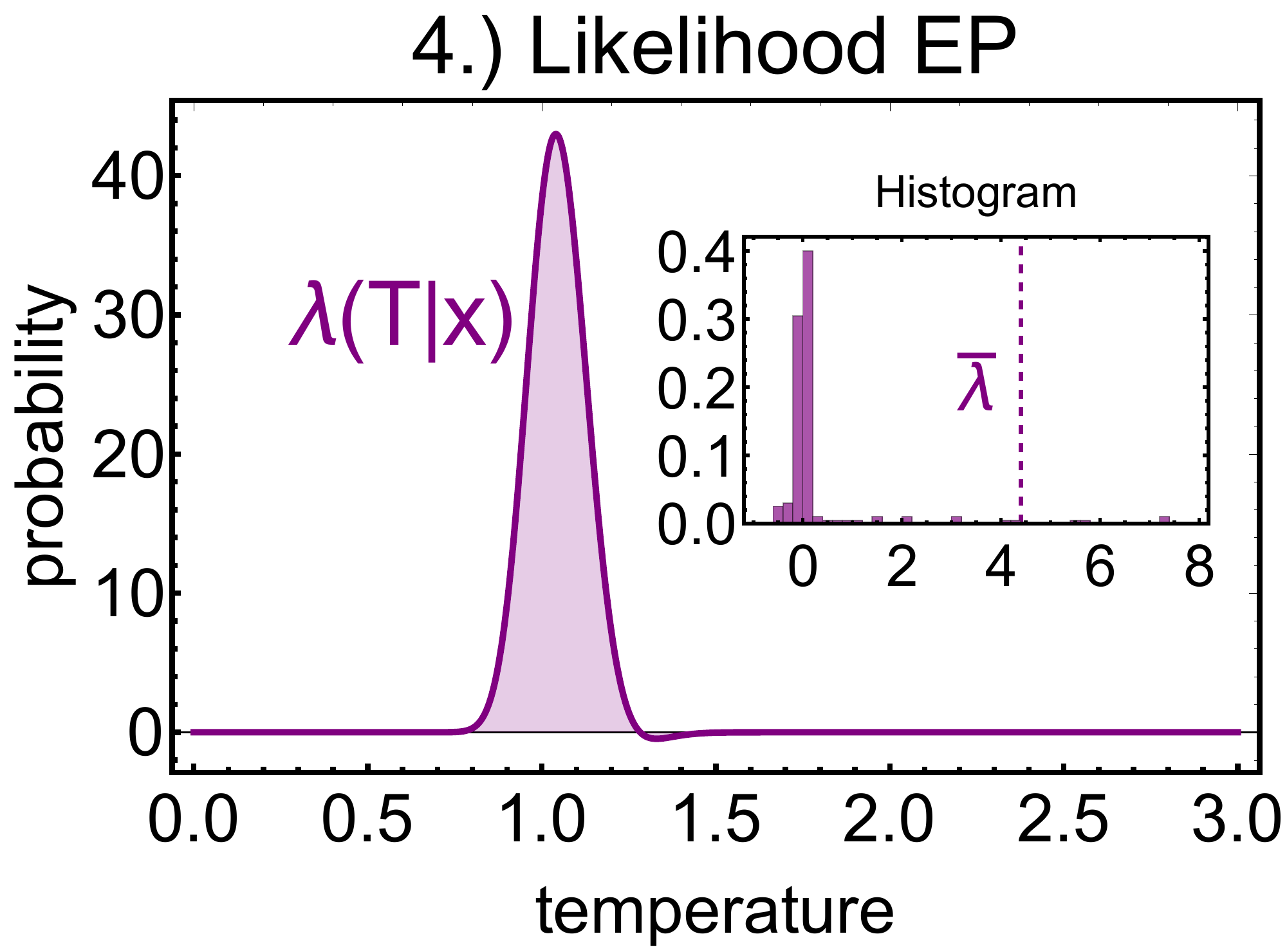}
\caption{Illustration of the second law of inference on a two-state system with uncertain temperature.
Before the experiment, the temperature is randomly chosen from the prior distribution $p(T)$ (top-left panel).
The experimenter cannot measure the temperature, but can observe the trajectory $\vx(t)$ (top-right panel).
From observing the trajectory, it is possible to update the temperature distribution $P(T) \mapsto P(T|\vx)$ (bottom-left panel).
Similarly, the temperature distribution obtained from observing the time-reversed trajectory under the time-reversed protocol is
$\tilde{P}(T|\tvx)$.
From their log ratio, it is possible to express the likelihood EP $\pmb{\lambda}(T|\vx)$ (bottom-right panel).
By calculating the histogram of $\lambda_\vx$, we observe that negative likelihood EP is observed much less common than positive
likelihood EP,
which corresponds to the detailed fluctuation theorem. Finally, the average value of likelihood EP remains positive,
which is in agreement with the second law of inference.}
\label{fig:traj}
\end{figure*}

To illustrate these results, consider a two-state system with states $\{0,1\}$, as depicted in \cref{fig:traj}. The
energy levels corresponding to the states are $E_0$ and $E_1$. We consider the transition rate matrix
\begin{equation}
K = \left(
         \begin{array}{cc}
           - e^{\frac{E_0-E_1}{T}} &  e^{\frac{E_1-E_0}{T}} \\
            e^{\frac{E_0-E_1}{T}} & - e^{\frac{E_1-E_0}{T}} \\
         \end{array}
       \right)
\end{equation}
The initial distribution is $p_0 = \{1/2,1/2\}$. Suppose that the temperature is randomly drawn from the Gamma distribution $P(T)
= T e^{-T}$ (top-left panel of \cref{fig:traj}).

The trajectory entropy production is
\begin{equation}
\pmb{\sigma}^T(\vx) = \ln p_{t_0}(\vx(t_0)) - \ln p(t_f)(\vx(t_f)) - \frac{1}{T} \left(E_{\vx(t_f)}-E_{\vx(t_0)}\right)
\end{equation}
By observing a trajectory $\vx$ (top-right panel of \cref{fig:traj}), we calculate an updated distribution $\vP(T|\vx)$ (bottom-right
panel of \cref{fig:traj}). Finally, we can also calculate the likelihood EP (bottom-right panel of \cref{fig:traj}). In the inset histogram of
the bottom-right panel, we see that the likelihood EP $\pmb{\lambda}_\vx$ can attain negative values, however $\bar{\lambda}_\vx =
\Lambda_\vx$ is greater than zero, which is in agreement with the detailed fluctuation theorem and the second law of inference.

\section{Discussion and Future work}

In any real-world experimental test of a system there are three major types of uncertainty:
a) uncertainty about the state of the system b) uncertainty about the state of the external environment that the system is
interacting
with, and c) uncertainty about the parameters of the equations governing the dynamics of the system and its interaction
with its external environment. In (classical) stochastic thermodynamics, the first type of uncertainty is addressed
by replacing the specification of the system's state (e.g., coarse-graining), and the second one is typically addressed by assuming the
environment
is infinite, at equilibrium, and evolving far faster than does the system (``separation of time scales''). In essence,
the entire field of stochastic thermodynamics concerns the consequences of those two types of uncertainty
for the dynamics of energy and particle counts in the system. However, very
little attention has been paid so far to the third type of uncertainty. Here we begin an investigation
of the consequences of that third type of uncertainty, showing that it entails major modifications to the
standard results previously derived in stochastic thermodynamics.

Our investigations have only scratched the surface of issues involved with this third type of uncertainty.
In the light of recent studies \cite{PhysRevLett.124.170601} where the system is coupled to a reservoir with fluctuating temperature, it makes sense to extend the analysis from a static distribution $p(\alpha)$ to a distribution over dynamic trajectories of the parameters, i.e., $\pmb{P}(\pmb{\alpha}_t)$. Another possible future direction can motivated by extending the framework to the case of stochastic thermodynamics hidden Markov models \cite{bechhoefer2015hidden} with the application to biophysical systems as flashing ratchets \cite{PhysRevLett.72.2652,PhysRevLett.105.150607} when the experimenter cannot directly observe whether the ratchet is on or off, and therefore does not also know the switching rate.

Some of the more immediate questions to be addressed in future work include the extension to the stochastic control protocols.
The first possible question is whether the \{quench; equilibrate; semi-statically-evolve\} protocol is the optimal protocol
which maps $p_x(t_i) \rightarrow p_x(t_f)$ or if there is an alternative protocol that generates less effective dissipated work.

Furthermore, it is reasonable to assume that in systems with limited possibility of measuring the system's parameters, it will be even more difficult to realize the desired control protocol with infinite precision. The related question is how the partial knowledge of the system obtained by the (imprecise) measurement from observing a trajectory, i.e., $P(\alpha) \mapsto P(\alpha|\vx_t)$ can be utilized to adjust the
control protocol  $\epsilon^{\star}(x) \mapsto \epsilon^{\star}(x|\vx_t)$.
In more realistic situations, the measurement cannot be done continuously but at given time instants and might bear some costs.
The question of optimal measurement and update of control protocol in uncertain environments is challenging but definitely a
crucial question
to be answered.

\begin{acknowledgements}
DHW would like to thank the Santa Fe Institute and the US National Science Foundation under grant 2221345 for support.  JK was supported by the Austrian Science Fund (FWF) project No. P 34994 and project No. P 33751 and Austrian Science Promotion Agency (FFG) Project Grant 857136.
\end{acknowledgements}

\bibliography{uncertain_thermo_refs.main}

\appendix

\section{Detailed calculation of moving optical tweezer with uncertain stiffness}
\label{app:tweezer}
Let us now show a detailed derivation of the moving optical tweezer with uncertain stiffness from the main text. Let us consider a particle described by an overdamped Langevin equation
\begin{equation}
 \dot{x} = - \mu \frac{\partial V}{\partial x} + \xi
\end{equation}
where $\xi(t)$ is the white noise, and $V$ is the potential. Let us consider the potential in the form
\begin{equation}
 V_k(x,t) = \frac{k}{2}(x-\lambda(t))^2
\end{equation}
where $k$ is the stiffness parameter and $\lambda(t)$ is the control protocol. The Sekimoto formula for the average work is
\begin{equation}
W[\lambda(t)] = \int_{t_i}^{t_f} \mathrm{d} t \dot{\lambda} \left\langle \frac{V_k(\lambda(t),x(t))}{\partial \lambda} \right\rangle
\end{equation}
where $\langle .. \rangle$ is the ensemble average. Let us consider $\mu=1$. By introducing $u(t) = \langle x(t) \rangle$, we obtain
\begin{equation}
\dot{u}_k = k(\lambda - u_k)
\end{equation}
where we omit the dependence of both $u$ and $\lambda$ on $t$. From this, we can express $\lambda$ as
\begin{equation}
  \lambda = \frac{\dot{u}_k}{k} + u_k
\end{equation}
Let us now express $\left\langle \frac{V_k(\lambda(t),x(t))}{\partial \lambda} \right\rangle$ as
\begin{equation}
\left\langle \frac{V_k(\lambda(t),x(t))}{\partial \lambda} \right\rangle = - k \langle x - \lambda \rangle = k (\lambda - u_k) = \dot{u}_k
\end{equation}
By plugging into the formula for work, we obtain
\begin{equation}\label{eq:w}
W[\lambda(t)] = \int_{t_i}^{t_f} \mathrm{d} t \left(\dot{u}_k^2 + \frac{\ddot{u}_k \dot{u}_k}{k} \right) =  \int_{t_i}^{t_f} \mathrm{d} t \dot{u}_k^2 + \frac{[\dot{u}_k^2]_{t_i}^{t_f}}{2k}
\end{equation}
Extremization of the work functional can obtained by $\frac{\delta W[\lambda(t)]}{\delta \lambda(t)} = 0$ which leads to the Euler-Lagrange equation
\begin{equation}
 \ddot{u}_k = 0
\end{equation}
which leads to $u(t) = m t$, where $m$ is a parameter to be determined. The boundary conditions are set to $u_k(t_i) = 0$, $\lambda(t_i) = 0$, from which we have $\dot{u}_k(t_i) = k(\lambda(t_i)-u_k(t_i)) = 0$. We set $\lambda(t_f) = \lambda_f$, thus, we have $\dot{u}_k(t_f) = k(\lambda_f - mt)$. Let us now set $t_i=0$
The total work is, therefore
\begin{equation}
W = m^2 t_f + \frac{k}{2} \left(\lambda_f - mt_f\right)^2
\end{equation}
The optimal solution is given by $m^\star = \frac{k \lambda_f}{2+k t_f}$. Thus, the optimal protocol can be expressed as
\begin{equation}
\lambda_k^\star = \frac{\lambda_f(1+kt)}{2+kt_f}
\end{equation}
and the optimal work is
\begin{equation}
W_k^\star = \frac{k \lambda_f^2}{2+kt_f}
\end{equation}
Note that we observe initial and final jumps in the protocol, i.e.,
\begin{eqnarray}
\Delta \lambda_k(t_i) = \lim_{t \rightarrow t_i^+} \lambda(t) - \lambda(t_i)
 \equiv \Delta \lambda_k(t_f) \nonumber\\
  = \lambda_k(t_f) - \lim_{t \rightarrow t_f^-}\lambda_k(t) = \frac{\lambda_f}{2+kt_f}
 \end{eqnarray}

Let us now focus on the case where $\lambda_k$ is used when the stiffness parameter is $\kappa$, which is not necessarily equal to $k$. In this case, we take the protocol $\lambda_k(t) = \frac{\lambda_f(1+kt)}{2 + k t_f}$ for $t_i \leq t \leq t_f$, and $\lambda(0) = 0$, $\lambda(t_f) = \lambda_f$, and plug it to the Langevin equation
$$\dot{u}_\kappa = \kappa(\lambda_k - u_\kappa)$$
together with the initial condition $u_\kappa(0) = 0$. By solving the differential equation, we get
\begin{eqnarray}
u_{\kappa}(t) = \frac{\lambda_f}{ (
   2+k t_f)} \left(k t + \left(1-\frac{k}{\kappa}\right)(1-e^{-\kappa t} )  \right)\nonumber\\
    = u_k(t) + \Delta_{k,\kappa}(t)
\end{eqnarray}
where $\Delta_{k,\kappa}(t)$ denotes the difference between $u_k$ and $u_\kappa$ for the same protocol $\lambda_k$. Note that $u_\kappa$ boils down to $u_k(t) = m^\star t$ for $k = \kappa$, as expected. The velocity is then
$$\dot{u}_\kappa(t) = \frac{\lambda_f}{ (
   2+k t_f)}\left(k + (\kappa - k)e^{-\kappa t}\right)$$
By plugging into the Eq. \eqref{eq:w}, we obtain that the work can be expressed as
\begin{widetext}
\begin{eqnarray}
W_\kappa[\lambda_k(t)] &=& \int_{0}^{t_f} \mathrm{d} t \dot{u}^2_\kappa(t) + \frac{[\dot{u}^2_\kappa]_{t_f}^{t_i}}{2 \kappa}\nonumber\\
&=& \int_{0}^{t_f} \mathrm{d} t \frac{\lambda_f^2}{ (
   2+k t_f)^2}\left(k + (\kappa - k)e^{-\kappa t}\right)^2
    + \frac{\kappa}{2} (\lambda_f - u_\kappa(t_f))^2  \nonumber\\
  &=& \frac{\lambda_f^2}{ (
   2+k t_f)^2} \left(k^2 t_f + \frac{2 k}{\kappa}(\kappa-k) (1-e^{-\kappa t_f})+\frac{(k-\kappa)^2}{2\kappa}(1-e^{-2 \kappa t_f}) \right)\nonumber\\
   &&+ \frac{\lambda_f^2 }{2(2+kt_f)^2 \kappa} \left( (k+\kappa)^2 +2 (k+\kappa)(\kappa-k) e^{-\kappa t_f} + (k-\kappa)^2 e^{-2 \kappa t_f} \right)\nonumber\\
&=& \frac{\lambda_f^2 }{(2+kt_f)^2}\left[ (2k  + k^2 t_f) + \frac{\kappa^2-k^2}{\kappa} + \frac{(k-\kappa)^2}{\kappa} e^{-\kappa t_f}\right]
\end{eqnarray}
Let us now consider that $\kappa$ is distributed by the distribution $p(\kappa)$. We choose our $k$ determining our protocol to be $k = \bar{\varkappa} = \int \mathrm{d} \varkappa \,  \varkappa \, p(\varkappa)$. Then the expected work (unadapted) $\bar{W} = \int \mathrm{d} \kappa W_\kappa(\lambda_{\bar{\kappa}}(t))$ is equal to
\begin{eqnarray}
W_{unad.} \equiv \bar{W} = \frac{\lambda_f^2 \bar{\kappa}}{(2+ \bar{\kappa}t_f)} +  \frac{\lambda_f^2 \bar{\kappa}}{(2+ \bar{\kappa}t_f)^2} \left(\bar{\kappa} - \bar{\kappa}^2 \int \mathrm{d} \kappa \frac{1} {\kappa} p(\kappa)\right) \nonumber\\
+
\frac{\lambda_f^2 }{(2+ \bar{\kappa}t_f)^2} \int \mathrm{d} \kappa \frac{(\bar{\kappa}-\kappa)^2}{\kappa} e^{-\kappa t_f} p(\kappa) = W_{\bar{\kappa}}[\lambda_{\bar{\kappa}}(t)]  + W_{diss}
\end{eqnarray}
\end{widetext}
On the other hand, if we were able to choose the optimal protocol for each $\kappa$, the expected work (adapted) would be
\begin{equation}
W_{ad.} = \overline{W_\kappa[\lambda_\kappa(t)]} = \int \mathrm{d} \kappa \frac{\lambda_f^2 \kappa}{2+\kappa t_f} p(\kappa)
\end{equation}
Since $\frac{\kappa}{2+\kappa t_f}$ is a concave function, we obtain from the Jensen inequality that
\begin{equation}
W_{ad.} \leq W_{unad.}
\end{equation}

\section{Brief review of stochastic thermodynamics}
\label{app:review_stoch_thermo}
In this section we present some additional
details of ordinary, no-uncertainty stochastic thermodynamics, and review some of the
main associated results.
We leave $N$ and associated parameters like the inverse temperatures $\beta^\nu$ and chemical potentials $\mu^\nu$ implicit.
We write the rate matrix for going from state $x'$ to state $x$ due to stochastic exchanges of
heat and / or particles with reservoir $\nu$ as $K_{x,x'}^\nu(t)$. From now on, we leave the time index implicit.
The full
rate matrix of the system is $K_{x,x'} = \sum_{\nu=1}^{N} K_{x,x'}^{\nu}$, and so
the master equation of the CTMC is $\dot{p}_t(x) = \sum_{x'} K_{x,x'} p_t(x')$.

Often in the literature, due to considerations involving time-symmetric microscale dynamics,
we assume that at any (implicit) time $t$
the separate matrices $K_{x,x'}^{\nu}$ each satisfy \emph{local detailed balance} (LDB) with respect to the energy level $u(x)$
at that time, for the particle reservoir $\nu$:

\begin{equation}
\label{eq:ldb}
\frac{K^{\nu} _{x,x'}}{K^{\nu}_{x'x}} = \frac{\pi^{\nu}_x}{\pi^{\nu}_{x'}}  = e^{-\beta^\nu[(u(x) - u(x'))
	 -\mu^\nu(n^\nu(x)-n^\nu(x'))]}\, .
\end{equation}
where $u(x)$ is the energy level of the system in state $x$ and $n^\nu(x)$ is the number of particles of the type specified by $\nu$
when the system is in state $x$. (If the reservoir $\nu$ does not exchange particles with the system, then $\mu^\nu = n^\nu(x) = 0$.)

%
Note that in the absence of chemical reservoirs, the map from rate matrices to energy functions is single-valued (up to an
overall additive constant), but the inverse map is multi-valued.


\subsubsection*{Ensemble thermodynamics}
The ensemble internal energy is written as \mbox{$U_t := \langle u_t \rangle = \sum_x p_t(x) u_t(x)$}.
The system exchanges particles with some of the reservoirs, as well exchanging energy with each of them directly
(e.g., via kinetic molecular collisions). \emph{The first law of thermodynamics} can be formulated as
\begin{equation}
\Delta U_t = \pmb{Q}_t + \pmb{W}_t  + \pmb{C}_t
\end{equation}
where $\Delta U_t := U_t - U_0$ and $\pmb{Q}_t := \int_0^t dt' \pmb{Q}_{t'}$ is total heat
flow into the system during the interval $[0, t]$, $\pmb{W}_t  := \int_0^t dt' \pmb{W}_{t'}$ is the total work
on the system during that interval, and $\pmb{C}_t
 := \int_0^t dt' \pmb{C}_{t'}$ is the total chemical work during that interval.
The heat flow rate $\dot{\pmb{Q}}_t$ can be decomposed into heat flows in
from the separate reservoirs, i.e., direct energy flows in from the separate reservoirs:
\eq{
\dot{\pmb{Q}}_t &= \sum_{\nu} \dot{\pmb{Q}}^\nu_t \nonumber\\
	&= \sum_{\nu} \sum_{xx'} K_{x,x'}^{\nu} p_t(x) (u_t(x)-\mu^\nu n^\nu(x))\, .
}
 For simplicity, we ignore the possibility of more than one distinguishable type of particle.
The mechanical work flow and chemical work flow are defined as
\eq{
\dot{\pmb{W}}_t   &= \sum_x  p_{t}(x) \dot{u}_t(x)\\
\dot{\pmb{C}}_t &= \sum_x \dot{p}_t(x) \sum_\nu \mu^\nu n^\nu(x)\, .
}

Shannon entropy at time $t$ is defined as \mbox{$S_t := \langle s_{t} \rangle = - \sum_x p_t(x) \ln p_t(x)$}.
The \emph{second law of thermodynamics} can be written as
\eq{
\Delta S_t = \pmb{\Sigma}_t + \pmb{\mathcal{E}}_t
}
where $\Delta S_t = S_t-S_0$, $\pmb{\Sigma}_t$ is the entropy production (EP) and $\pmb{\mathcal{E}}_t$ is the entropy flow (EF).
%
The {EF} rate is
\eq{
\dot{\pmb{\mathcal{E}}}_t =  -\sum_\nu  \sum_{x,x'} K^\nu_{x,x'} p(x') \ln \dfrac{K^\nu_{x,x'}}{K^\nu_{x',x}}
}
so that the EP rate is
\eq{
\dot{\pmb{\Sigma}}_t =  -\sum_\nu  \sum_{x,x'} K^\nu_{x,x'} p(x') \ln \dfrac{K^\nu_{x,x'}p_t(x')}{K^\nu_{x',x}p_t(x)}\, .
\label{eq:6}
}
The second law of thermodynamics is enforced by the fact that
for any rate matrix, EP rate is non-negative, i.e., $ \dot{\pmb{\Sigma}}_t \geq 0$.
When LDB holds, the EF rate can be expressed in terms of \emph{thermodynamic entropy}, i.e.,
\eq{
\dot{\pmb{\mathcal{E}}}_t =  \sum_\nu \beta^\nu \dot{\pmb{Q}}^{\nu}_t\, .
}



\subsubsection*{Trajectory thermodynamics}
The {\emph{trajectory internal energy}} is written as $\pmb{u}_{t}(\vx)$.
The first law of thermodynamics on the trajectory level for any
time $t$ is
\begin{equation}
\frac{\ud}{\ud t} \pmb{u}_t(\vx) = \dot{\pmb{q}}_t(\vx) + \dot{\pmb{w}}_t(\vx) + \dot{\pmb{c}}_t(\vx)
\end{equation}
where
\eq{
\dot{\pmb{q}}_t(\vx) &= \sum_\nu \dot{\pmb{q}}_t^\nu(\vx)\nonumber\\
&= \sum_{\nu} \sum_x  \dot{\delta}_{x,\vx(t)} (u_t({x})-\mu^\nu n^\nu(x)), \\
\dot{\pmb{w}}_t(\vx) &= \sum_x \delta_{x,\vx(t)} \dot{u}_t{(x)}, \\
\dot{\pmb{c}}_t(\vx) &= \sum_{\nu} \sum_x  \dot{\delta}_{x,\vx(t)} \mu^\nu n^\nu(x)
\label{eq:4}
}
are called the trajectory heat, trajectory mechanical work and trajectory chemical work, respectively.

Trajectory entropy is defined as $s_t(\vx) := - \ln p_t(\vx(t))$. Then time derivative of entropy can be decomposed as
\eq{
\frac{\ud}{\ud t} \pmb{s}_t(\vx) &= \dot{\pmb{\sigma}}_t(\vx) + \dot{\pmb{\epsilon}}_t(\vx)
\label{eq:stoch_entr_fixed_rate_matrix}
}
where due to LDB,  \emph{trajectory EF rate} is
\eq{
\dot{\pmb{\epsilon}}(\vx) =  \sum_\nu  \beta_\nu \dot{\pmb{q}}^\nu(\vx) = \sum_x \dot{\delta}_{x,\vx(t)} \sum_\nu \ln
\dfrac{K^\nu_{x,x'}}{K^\nu_{x',x}}\, .
\label{eq:12}
}

ensemble-level versions, i.e., $\boldlangle {\dot{\pmb{q}}}^\nu_t \boldrangle = \dot{\pmb{Q}}^\nu_t$, $\boldlangle
\dot{\pmb{\epsilon}_t} \boldrangle = \dot{\pmb{\mathcal{E}}}_t$,

\section{Proof that the $H^*$ solving \cref{eq:3} is unique}
\label{app:proof_of_unique_H*}

Write the (countable) elements of $X$ as $1, 2, \ldots$,  ordered
so that $\overline{p}_1(t_i) \ge \overline{p}_2(t_i) \ge \ldots$. For simplicity, we assume that $\overline{p}_{|X|}(t_i) > 0$,
i.e., we assume that all elements of $X$ have nonzero probability under $\overline{p}(t_i)$; it is straightforward to extend the
analysis below
to the case where $\overline{p}_x(t_i) = 0$ for some $x$, by setting $H^*(x) = \infty$ for any such $x$.

By inspection, if there is a solution $H^*(x) = H(x)$ for some given $\overline{p}(t_i)$ and $\ud P_\alpha$, then
$H(x) + k$ is also a solution for that $\overline{p}(t_i)$ and $\ud P_\alpha$, for any real number $k$. Accordingly, wolog set
$H^*(1) = 1$.
For use below, write the maximal element of $\supp (\ud P_{\beta^\alpha})$ as $\beta_{max}$.

The proof is by iterative construction of $H^*(x)$. To begin, we set $H^*(i) = i$ for all $i$.
We then run a two-step iterative procedure, from $i = 2$ to $i = |X|$, changing each element $H^*(i)$ in turn
so that
\eq{
\dfrac{\int \ud P^\alpha p^{eq}_{H^*,{\beta^\alpha}}(i-1)}{\int \ud P^\alpha p^{eq}_{H^*,{\beta^\alpha}}(i)} &=
		\dfrac{\overline{p}_{i-1}(t_i)}{\overline{p}_i(t_i)}
\label{eq:3+}
}
When this iterative procedure finishes, we will know that \cref{eq:3+} is satisfied for all $i \ge 2$ for the finishing $H^*$,
and so \cref{eq:3} is met for that $H^*$.

In the first step of the iterative procedure, if $\overline{p}_i(t_i) = \overline{p}_{i-1}(t_i)$, we set $H^*(i) = H^*(i-1)$.
This guarantees that \cref{eq:3+} holds for this particular $i$.

If instead $\overline{p}_i(t_i) < \overline{p}_{i-1}(t_i)$ for this $i$, then we execute the second step of the
iterative procedure. Consider two candidate Hamiltonians, $H^+(x)$ and $H^-(x)$, which both equal
the current $H^*(j)$ for all $j \ne i$. We complete their definitions by setting $H^+(i) = H^*(i)$ and $H^-(i) = \kappa$, where
\eq{
\kappa > H^-(i) + \beta_{max}^{-1}\ln\dfrac{\overline{p}_{i-1}(t_i)}{\overline{p}_i(t_i)}
}
This guarantees that
\eq{
\dfrac{\overline{p}_{i-1}(t_i)}{\overline{p}_i(t_i)} &< \dfrac{e^{-\beta_{max} H^-(i-1)}}{e^{-\beta_{max} H^-(i)}}
}

Plugging in these definitions establishes both that
\eq{
\dfrac{\int \ud P^\alpha p^{eq}_{H^+,{\beta^\alpha}}(i-1)}{\int \ud P^\alpha p^{eq}_{H^+,{\beta^\alpha}}(i)} &<
		\dfrac{\overline{p}_{i-1}(t_i)}{\overline{p}_i(t_i)}
\label{eq:102a}
}
and
\eq{
\dfrac{\int \ud P^\alpha p^{eq}_{H^-,{\beta^\alpha}}(i-1)}{\int \ud P^\alpha p^{eq}_{H^-,{\beta^\alpha}}(i)} &>
		\dfrac{\overline{p}_{i-1}(t_i)}{\overline{p}_i(t_i)}
\label{eq:103a}
}
$\int \ud P^\alpha p^{eq}_{H^*,{\beta^\alpha}}(i)$ is a differentiable function of $H^*(i)$,
and $H^+(j) = H^-(j) = H^*(j)$ for all $j \ne i$. Therefore applying the intermediate
value theorem with \cref{eq:102a} and \cref{eq:103a} means that there is some value $H^*(i)$ such that \cref{eq:3+} holds.
The second step of the iterative procedure finishes by setting $H^*(i)$ to this intermediate value. At that point we
increment $i$ by $1$.

At the end of the second step for any particular $i$, we have established that \cref{eq:3+} holds for that $i$ in both of the
possible
situations  $\overline{p}_i(t_i) = \overline{p}_{i-1}(t_i)$ and $\overline{p}_i(t_i) < \overline{p}_{i-1}(t_i)$. Therefore
by the end of the iterative procedure we have established that \cref{eq:3+} holds for all $i$.

This completes the proof that there is a solution for $H^*(x)$, as claimed. Uniqueness of this solution
follows from the fact that the derivative of  $\int \ud P^\alpha p^{eq}_{H^*,{\beta^\alpha}}(i)$ with respect
to $H^*(i)$ is non-negative, and so the intermediate value arising in the second step of
the iterative procedure is unique.

\section{Proof of non-increasing value of information for no-uncertainty rate matrices}
\label{app:value_of_i}
Consider the case where the rate matrix $K_{xx'}(t)$ during the interval $t \in [t_i, \tau)$ has
no uncertainty. We can see how the thermodynamic value of information of $\alpha$ depends on $\tau$ in this case,
by taking the derivative of the LHS of $I_\tau$ with respect to $\tau$:
\begin{widetext}
\eq{
\dfrac{d  I_{\tau}}{d\tau} &= \dfrac{\mathrm{d} S\left(\overline{p}_\tau(X)\right)}{\mathrm{d}\tau} -  \int \ud P^\alpha
\dfrac{\mathrm{d} S\left(p_\tau^\alpha\right)}{\mathrm{d}\tau} \nonumber\\
	&= \int \ud P^\alpha \,  \sum_{x,x'} K_{xx'}(\tau) p^\alpha_\tau(x') \left[ \ln \dfrac{\int \ud P^{\alpha'} K_{xx'}(\tau)
			p_\tau^{\alpha'}(x')}{\int \ud P^{\alpha'}  K_{x'x}(\tau) p_\tau^{\alpha'}(x)}
					- \ln \dfrac{K_{xx'}(\tau) p_\tau^\alpha(x')}{K_{x'x}(\tau) p_\tau^\alpha(x)} \right]
\label{eq:10}
}
Therefore the time-derivative of the value of information reduces to a difference of EP rates,
distinguished from each other by whether we know $\alpha$ or not. In addition, if we multiply and divide by $P_\alpha$ inside the
rightmost logarithm in \cref{eq:10}
(i.e., change the two conditional probability distributions into joint probability densities), and consider $\mathrm{d} P^\alpha =
\mathrm{d} \alpha P^\alpha$ we get
\eq{
\dfrac{d  I_{\tau}}{d\tau} &= \sum_{x,x'} \int \ud \alpha P^\alpha \,  K_{xx'}(\tau)  p^\alpha_\tau(x') \left[ \ln \dfrac{\int \ud
\alpha P^{\alpha'} K_{xx'}(\tau)
			p_\tau^{\alpha'}(x')}{\int \ud \alpha  P^{\alpha'}  K_{x'x}(\tau) p_\tau^{\alpha'}(x)}
					- \ln \dfrac{P^\alpha K_{xx'}(\tau) p_\tau^\alpha(x')}
									{P^\alpha K_{x'x}(\tau) p_\tau^\alpha(x)} \right]
\label{eq:105}
}
In particular, if $P_\alpha$ is a probability mass function, $p(\alpha)$, then \cref{eq:105} can be written in terms of the joint
probability $p_\tau(x,\alpha) = p(\alpha) p^\alpha_\tau(x)$:
\eq{
\sum_{x,x'}  \sum_\alpha    K_{xx'}(\tau)  p_\tau(x',\alpha) \left[ \ln \dfrac{ \sum_{\alpha'}   K_{xx'}(\tau)
			p_\tau(x',\alpha')}{\sum_{\alpha'}  K_{x'x}(\tau) p_\tau(x, \alpha')}
					- \ln \dfrac{K_{xx'}(\tau) p_\tau(x', \alpha)}
									{K_{x'x}(\tau) p_\tau(x,  \alpha)} \right]
\label{eq:104}
}
\end{widetext}

Each $(x, x')$ pair in the outer sum in \cref{eq:104} in which $x = x'$ gives a value of zero, since both of the logarithms
equal zero if $x' = x$. In addition, $K^{xx'}(\tau)$ is non-negative for all $x' \ne x$.
So we can apply the log-sum inequality (assuming $a_i$ and $b_i$ non-negative):
\begin{equation}
\sum_i a_i \log \frac{a_i}{b_i} \geq \left(\sum_i a_i\right) \log \frac{\sum_i a_i}{\sum_i b_i}
\end{equation}
separately for each $(x, x' \ne x)$ pair. This means that $\frac{\mathrm{d} I_\tau}{\mathrm{d} \tau} \leq 0$,
as claimed.

\end{document}